\documentclass[12pt,a4paper]{article}
\usepackage{amsmath,amssymb,amsfonts}
\usepackage{graphicx,epsfig,url}
\usepackage{color}
\usepackage{amsmath}
\usepackage{amsfonts}
\usepackage{amssymb}
\usepackage{float}
\usepackage{caption2}
\usepackage{graphicx}%
\numberwithin{equation}{section} 
\setcaptionmargin{1cm}
\def\beq{\begin{equation}}
\def\eeq{\end{equation}}
\def\bea{\begin{eqnarray}}
\def\eea{\end{eqnarray}}
\def\bitem{\begin{itemize}}
\def\eitem{\end{itemize}}
\def\bar#1{\overline{#1}}

\def\abs#1{\left| #1\right|}

\def\inv{^{\raise.15ex\hbox{${\scriptscriptstyle -}$}\kern-.05em 1}}
\def\lbar{{\lower.35ex\hbox{$\mathchar'26$}\mkern-10mu\lambda}} 

\def\to{\rightarrow}

\let\<=\langle
\let\>=\rangle

\let\+=\uparrow
\let\al=\alpha

\let\ga=\gamma
\let\Ga=\Gamma
\let\de=\delta

\let\la=\lambda

\let\Om=\Omega

\begin{document}
\begin{titlepage}
\begin{flushright}
OUTP-09-18-P, UCB-PTH-09/32
\end{flushright}
\vskip 1.0cm
\begin{center}

{\Large \bf Freeze-In Production of FIMP Dark Matter} \vskip 1.0cm {\large 
Lawrence J. Hall$^a$,\ \   Karsten Jedamzik$^b$, \\[10pt]
John March-Russell$^c$ and Stephen M. West$^{d,e}$} \\[1cm]
{\it $^a$ Department of Physics, University of California, Berkeley and\\
Theoretical Physics Group, LBNL, Berkeley, CA 94720, USA, and \\
Institute for the Physics and Mathematics of the Universe, \\ University of Tokyo, Kashiwa 277-8568, Japan}\\
{\it $^b$ Laboratoire de Physique Th«eorique et Astroparticules, UMR5207-CNRS,\\
UniversitÕe Montpellier II, F-34095 Montpellier, France} \\
{\it $^c$ Rudolf Peierls Centre for Theoretical Physics, University of Oxford,\\
1 Keble Rd., Oxford OX1 3NP, UK}\\
{\it $^d$ Royal Holloway, University of London, Egham, TW20 0EX, UK} \\
{\it $^e$ Rutherford Appleton Laboratory, Chilton, Didcot, OX11 0QX, UK} \vskip 1.0cm 
\vspace{-0.8cm}
(December 8th, 2009)\\

\abstract{We propose an alternate, calculable mechanism of dark matter
genesis, ``thermal freeze-in," involving a Feebly Interacting Massive Particle (FIMP)
interacting so feebly with the thermal bath that it never attains thermal 
equilibrium.   As with the conventional ``thermal freeze-out" production mechanism, 
the relic abundance reflects a combination of initial thermal distributions together with 
particle masses and couplings that can be measured in the laboratory or astrophysically.  
The freeze-in yield is IR dominated by low temperatures near the FIMP mass and is 
independent of unknown UV physics, such as the reheat temperature after inflation.  
Moduli and modulinos of string theory compactifications that receive mass from 
weak-scale supersymmetry breaking provide implementations of the freeze-in 
mechanism, as do models that employ Dirac neutrino masses or 
GUT-scale-suppressed interactions. Experimental signals of freeze-in and FIMPs
can be spectacular, including the production of new metastable coloured
or charged particles at the LHC as well as the alteration of big bang
nucleosynthesis.  
}
\end{center}
\end{titlepage}

\tableofcontents

\section{Introduction}

\begin{figure}
\vspace{-2cm}
\centerline{\includegraphics[width=14cm]{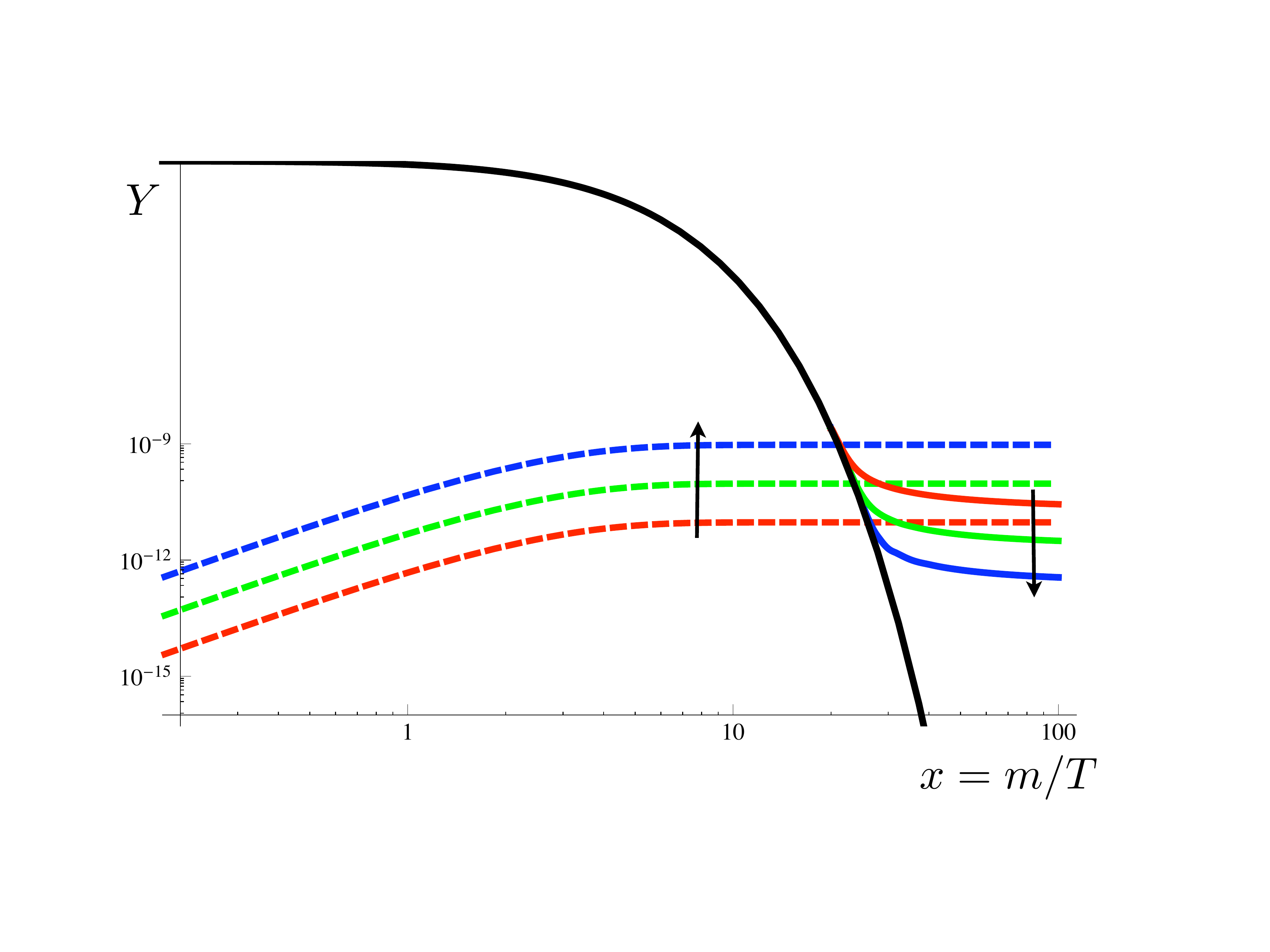}}
\vspace{-1.5cm}
\caption{Log-Log plot of the evolution of the relic yields for conventional freeze-out (solid coloured) and freeze-in via a Yukawa interaction (dashed coloured) as a function of $x=m/T$.  The black solid line indicates the yield assuming equilibrium is maintained, while the arrows indicate the effect of increasing coupling strength for the two processes.  Note that the freeze-in yield is dominated by the epoch $x\sim 2-5$, in contrast to freeze-out which only departs from equilibrium for $x\sim 20-30$.\label{fi_vs_fo}}
\end{figure}

Many theories of Dark Matter (DM) genesis are based upon the mechanism of ``thermal freeze out" \cite{freezeout}.  In this process DM particles have a large initial thermal density which, as the temperature of the hot plasma of the early universe drops below the mass, dilutes away until the annihilation to lighter species becomes slower than the expansion rate of the universe and the comoving number density of DM particles becomes fixed.
The larger this annihilation cross section the more the DM particles are able to annihilate and hence a thermal distribution with an exponential Boltzmann factor is maintained to a lower temperature, giving a lower final yield.  An attractive feature of the freeze-out mechanism is that for renormalisable couplings the yield is dominated by low temperatures with freeze-out typically occurring at a temperature a factor of $20-25$ below the DM mass, and so is independent of the uncertain early thermal history of the universe and possible new interactions at high scales. 

Are there other possibilities, apart from freeze-out, where a relic abundance reflects a combination of initial thermal distributions together with particle masses and couplings that can be measured in the laboratory or astrophysically? In particular we seek cases, like the most attractive form of freeze-out, where production is IR dominated by low temperatures of order the DM mass, $m$, and is independent of unknown UV quantities, such as the reheat temperature after inflation. 

In this paper we show that there is an alternate mechanism, ``freeze-in", with these features.   Suppose that at temperature
$T$ there is a set of bath particles that are in thermal equilibrium and some other long-lived particle $X$, having interactions with the bath that are so feeble that $X$ is thermally decoupled from the plasma.  We make the crucial assumption that the earlier history of the universe makes the abundance of $X$ negligibly small, whether by inflation or some other mechanism.  Although feeble, the interactions with the bath do lead to some $X$ production and, for renormalisable interactions, the dominant production of $X$ occurs as $T$ drops below the mass of $X$ (providing $X$ is heavier than the bath particles with which it interacts). The abundance of $X$ ``freezes-in" with a yield that increases with the interaction strength of $X$ with the bath.

Freeze-in can be viewed as the opposite process to freeze-out.   
As the temperature drops below the mass of the relevant particle, the DM 
is either heading away from (freeze-out) or towards (freeze-in) thermal 
equilibrium.  Freeze-out begins with a full $T^3$ thermal number density 
of DM particles, and reducing the interaction strength helps to maintain 
this large abundance.  Freeze-in has a negligible initial DM abundance, but 
increasing the interaction strength increases the production from the thermal 
bath. These trends are illustrated in 
Figure~\ref{fi_vs_fo}, which shows the evolution with temperature of the dark matter abundance
according to, respectively, conventional freeze-out, and the freeze-in mechanism
we study here.

In Section \ref{idea}, as well as outlining the basic freeze-in mechanism and comparing its features with those of freeze-out,
we also introduce the idea of a FIMP---a ``Feebly-Interacting-Massive-Particle" (or alternatively ``Frozen-In-Massive-Particle")---as distinct from a WIMP whose relic abundance is set by conventional freeze-out.  Two cases are considered: either the FIMP itself is the dark matter which is frozen-in, or the dominant contribution to the DM density arises from frozen-in FIMPs which then decay to a lighter DM particle.  For enhanced pedagogy the detailed calculation of the DM abundance in these cases is postponed until Section \ref{calculation}. 

We turn in Section \ref{candidates} to the question of motivated DM candidates that
have a relic abundance determined by the IR-dominated freeze-in mechanism, and show that the moduli
and modulinos of compactified string theories with weak-scale supersymmetry breaking 
provide implementations of the freeze-in mechanism, as does any extra-dimensional extension 
of the Standard Model (SM) with some moduli stabilised at the weak scale.   Models following
from Dirac neutrino masses, or involving kinetic mixing with a hidden sector also naturally
accommodate FIMPs and the freeze-in mechanism.  
A striking difference with freeze-out production of DM, is that for freeze-in the final relic density
is, in the simplest cases, automatically independent of the FIMP mass, allowing superheavy
as well as weak-scale mass DM candidates, and we touch on a high-scale extra-dimensional
realisation of such a scenario.  Theories utilising GUT-scale-suppressed non-renormalisable
interactions involving SM or MSSM Higgs fields which become renormalisable when the Higgs gets its vacuum
expectation value also naturally give rise to FIMPs.  However, in this case, there is also a UV contribution
to the freeze-in yield which can dominate the IR contribution if the reheat temperature is sufficiently large, and
we therefore postpone the discussion of such implementations until Section \ref{comments}  where the effect of
higher-dimension-operators on freeze-in yields is considered.

Experimental signals of freeze-in and FIMPs are briefly summarised in Section 
\ref{signatures}, with a companion paper \cite{companion} more extensively discussing
the rich phenomenology that accompanies freeze-in to follow. The signals
depend on the nature of the Lightest Observable 
Sector Particle (LOSP) that carries the conserved quantum number that 
stabilises DM.  (In theories with weak scale supersymmetry, the LOSP is the 
lightest superpartner in thermal equilibrium at the weak scale.)  There are 
two general possibilities:  The DM particle is the FIMP, and there are 
spectacular LHC signatures arising from the production of coloured or charged 
LOSPs, which stop in the detector and decay with a long lifetime.
In addition, LOSPs decaying to FIMPs in the early Universe may lead to
interesting modifications of Big-Bang Nucleosynthesis (BBN).   
Alternatively, the DM particle is the LOSP but its interactions are too strong to have a sufficiently large freeze-out relic density; instead its relic abundance results from freeze-in of FIMPs, with the FIMPs later decaying to the DM.  The second scenario can give signals via alteration of big-bang nucleosynthesis element abundances, and via increased DM pair annihilation relevant for indirect detection of DM.

Thermal relic abundances conventionally arise by decoupling of a species that was previously in thermal equilibrium, whether with or without a chemical potential.  Freeze-in provides the only possible alternative thermal production mechanism that is dominated by IR processes.  In Section \ref{phasediagrams} we sketch ``abundance phase diagrams" that show regions of mass and coupling where each mechanism dominates the production of the relic abundance. The topology of these phase diagrams, as well as the number of domains where
different mechanisms dominate, depends on the form of the DM-bath interaction, as we illustrate with two examples arising from a quartic scalar interaction and a Yukawa interaction.  Furthermore, we argue that there exists a certain ``universality" to the phase diagram behaviour at small coupling.   

After presenting the detailed calculation of the relic abundance in various cases in Section~\ref{calculation}, we discuss the physics of FIMP interactions mediated by higher-dimension-operators, as well as some variations of the basic freeze-in mechanism in Section \ref{comments}.  Finally we conclude in Section \ref{conclusions}.

\section{General Features of Freeze-In}\label{idea}

The basic mechanism of freeze-in is simple to describe although, as we will argue later,
there can be many variations with important differences of detail and signals.  Here we 
give the general mechanism.  At temperatures well above the weak scale we assume that there 
is a FIMP, $X$, that is only very weakly coupled to the thermal bath via some renormalisable 
interaction.   The interaction vertex may involve more than one particle from the thermal bath and 
the mass of the heaviest particle at the vertex is $m$, which we typically 
take to be near the weak scale.  For a Yukawa or quartic interaction, the dimensionless coupling 
strength is $\lambda$, while for a trilinear scalar interaction the coupling is $\lambda m$, 
and in all cases $\lambda \ll 1$.  

At very high temperatures we assume a negligible initial $X$ abundance.  
As the universe evolves $X$ particles are produced from collisions or decays of bath
particles, but at a rate that is always suppressed by $\lambda^2$.  During a Hubble doubling 
time at era $T \gg m$, the $X$ yield is
\begin{equation}
Y(T)  \, \sim \, \lambda^2 \, \frac{M_{Pl}}{T} \, \left( 1, \, \frac{m^2}{T^2} \right),
\label{eq:YT}
\end{equation}
where $Y=n/S$, $n$ is the number density of $X$, and $S$ is the entropy density of the plasma. 
The $m$ independent yield corresponds to quartic interactions, while the additional 
$m^2/T^2$ suppression applies to Yukawa interactions
and the super-renormalisable case.  The process is always 
IR dominated, favouring low temperatures.  The dominant production occurs at $T \sim m$, 
since at lower temperatures there will be an exponential suppression resulting from the 
necessity of involving a particle of mass $m>T$.  Hence, for all renormalisable interactions, 
the abundance of $X$ ``freezes-in" with a yield
\begin{equation}
Y_{FI} \, \sim \lambda^2 \left( \frac{M_{Pl}}{m} \right).
\label{eq:YFI}
\end{equation}
%

\begin{figure}
\vspace{-1cm}
\centerline{\includegraphics[width=13cm]{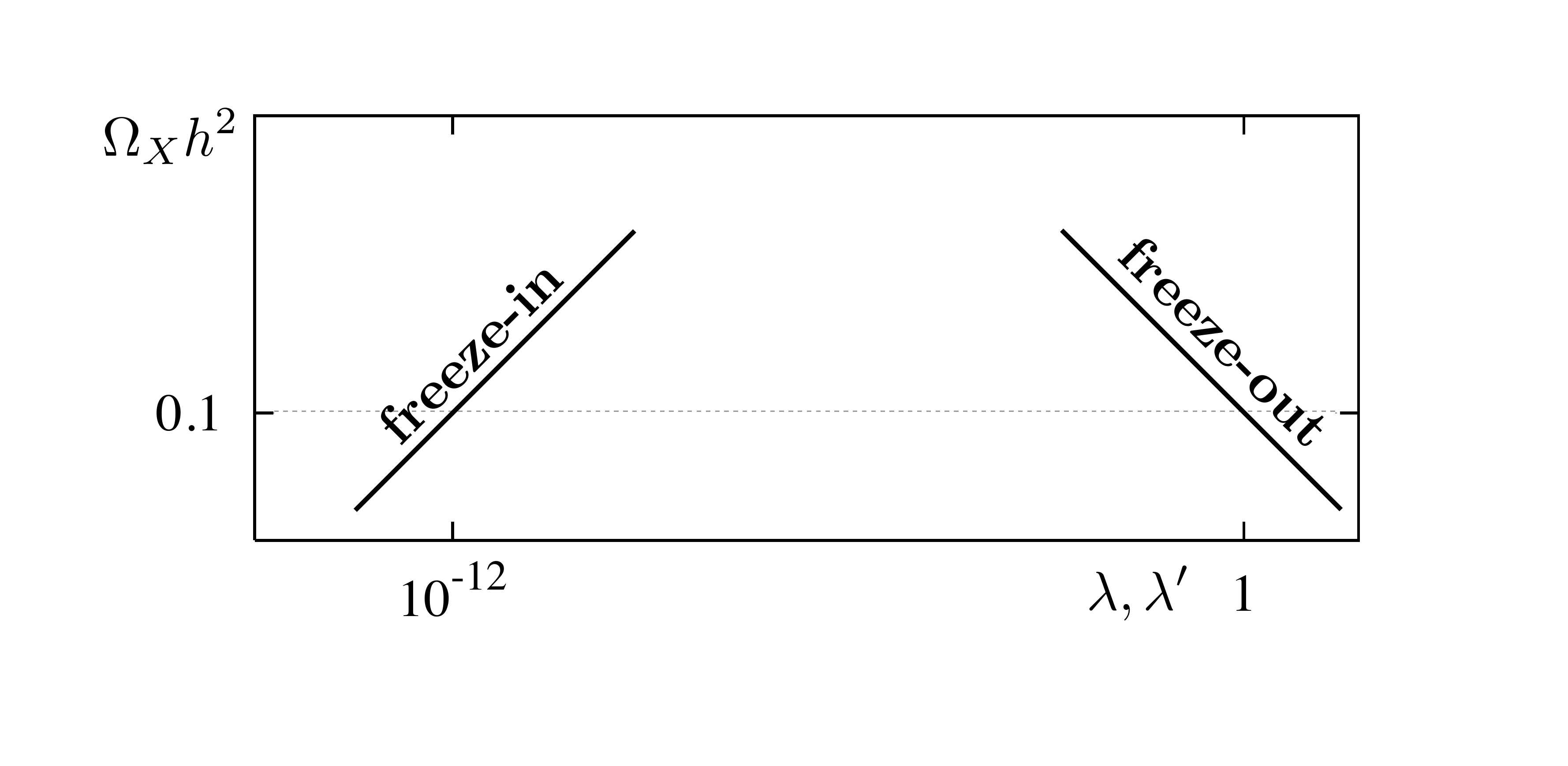}}
\vspace{-1cm}
\caption{Schematic picture of the relic abundances due to freeze-in and freeze-out as a function of coupling strength. The way in which the freeze-out and freeze-in yield behaviours connect to one another is model-dependent.  As we show in detail in Section \ref{phasediagrams}, freeze-in and freeze-out are in fact two of the four basic mechanisms for thermal DM production, and we sketch the ``abundance phase diagrams" of DM depending upon the strength and type of the DM-thermal bath interaction and the DM mass.\label{simple}}
\end{figure}

As mentioned in the Introduction, freeze-in can be viewed as the opposite process to freeze-out.  We recall that, in the absence of a chemical potential, the freeze-out yield is given by
\begin{equation}
Y_{FO} \sim \frac{1}{\sigma v  M_{Pl} \, m'}. 
\label{eq:Yfo}
\end{equation}
In the simple case that the DM mass $m'$ is the only mass scale entering $\sigma v$, we have $\sigma v \sim \lambda'^2/m'^2$, where $\lambda'$ is the relevant interaction strength, giving a freeze-out yield of
\begin{equation}
Y_{FO} \sim \frac{1}{\lambda'^2} \left( \frac{m'}{M_{Pl}} \right).
\label{eq:YFO}
\end{equation}
Freeze-in assumes a negligible initial $X$ abundance, but increasing the interaction strength, $\lambda$, increases the production from the thermal bath, while freeze-out begins with a full $T^3$ thermal number density of DM particles, and reducing the interaction strength, $\lambda'$, helps to maintain this large abundance (see Figure \ref{simple}). 
Indeed, the yields eqns.(\ref{eq:YFI}) and (\ref{eq:YFO}) show inverse dependences on the coupling and mass, which is
stressed by writing  
\begin{equation}
Y_{FI} \, \sim \lambda^2 \, m t_m,  \hspace{1in} Y_{FO} \, \sim \frac{1}{\lambda'^2 \, m' t_{m'}},
\label{eq:Y}
\end{equation}
where $t_m \sim M_{Pl}/m^2$ ($t_m' \sim M_{Pl}/m'^2$) is the 
Hubble time at the epoch of freeze-in (freeze-out).  The freeze-out abundance decreases with $t_{m'}$ while the freeze-in abundance is increased by occurring at late times.

Despite these opposite features, freeze-out and freeze-in share crucial common aspects: the final out-of-equilibrium abundance, given the relevant particle masses and couplings, can be computed {\it solely from an initial state of bath particles that are in thermal equilibrium}, and the resulting abundance is {\it dominated by IR physics}.  

For freeze-out the special case $\lambda' \sim 1$ and $m' \sim v$, the scale of weak interactions, 
gives DM as  ``Weakly Interacting Massive Particles", or WIMPs,
with 
\beq
Y_{FO} \sim \frac{v}{M_{Pl}}.
\eeq
In practice the cross section may involve more than one mass scale in the TeV domain, so that there are orders of magnitude spread in the abundance expected from WIMP dark matter.  Nevertheless, the parametric understanding of the abundance of DM is suggestive, and the prediction of a TeV mass particle with coupling strengths of order unity offers the hope of collider verification of the production mechanism. 
For freeze-in, if the particle masses are again at the weak scale, the same dependence of the relic abundance on $v$ and $M_{Pl}$ follows if the small coupling $\lambda$ is linear in the weak scale:
\beq
\lambda \sim \frac{v}{M_{Pl}} \hspace{0.3in} \rightarrow \hspace{0.3in} Y_{FI} \sim  \frac{v}{M_{Pl}}.
\eeq
As in the WIMP case, this parametric behaviour is significantly modified by numerical factors;
nevertheless, it suggests seeking theories where small couplings arise at linear order in the 
weak scale. 

Whether produced by freeze-out or freeze-in, stable DM is the lightest particle transforming 
non-trivially under some unbroken symmetry.   For conventional freeze-out, this lightest particle is
automatically the WIMP, whereas for freeze-in two particles are of interest: the FIMP 
and the lightest particle in the thermal bath that carries the symmetry, the LOSP.  
If the LOSP is lighter, then LOSP DM is produced by FIMP decay.  If FIMPs are lighter, 
then collider signals involve the production of LOSPs followed by decays to FIMPs.  In either case,
the freeze-in mechanism always introduces particles of very long lifetime
\beq
\Gamma_{FIMP/LOSP} \sim \lambda^2 m \sim \frac{v^3}{M_{Pl}^2}
\eeq
where $m$ is the mass of the decaying FIMP or LOSP, and in the last expression we took 
$m\sim v$ and $\lambda \sim v/M_{Pl}$.   The heavier of the LOSP and FIMP will 
generically decay to the lighter with a lifetime within a few orders of magnitude of a second. 

We show in detail in Section \ref{calculation} that the freeze-in density is dominated, 
where possible, by decays or inverse decays involving the bath particles and $X$.  
Freeze-in of a stable $X$ via the decays of LOSPs in the thermal bath (see eq.~(\ref{eq:om12}) and surrounding discussion) gives
\beq
\Om_X h^2=\frac{1.09\times 10^{27}}{g^S_* \sqrt{g^{\rho}_*}} \frac{m_X \Ga_{B_1}}{m_{B_1}^2},
\eeq
where $B_1$ is the LOSP.  For a LOSP decay rate given by
$\Ga_{B_1}=\lambda^2 m_{B_1}/8\pi$ and, for simplicity,
making the good approximation $g^S_* \simeq g^{\rho}_*$, the required DM
density $\Omega_X h^2 \simeq 0.106$ occurs for a coupling of size
\beq
\lambda \simeq 1.5\times 10^{-12} \left(\frac{m_{B_1}}{m_X}\right)^{1/2}  \left(\frac{g_*(m_{B_1})}{10^2}\right)^{3/4}. 
\eeq
As discussed in Section~\ref{calculation} this value is reduced to $\lambda \sim 1.5\times10^{-13} (10^2 /g_{bath})^{1/2}$ if
the FIMP couples with comparable strength to many ($g_{bath}$) bath particles of comparable mass, as occurs, for example,
in supersymmetric realizations of FIMPs as discussed in Section~\ref{candidates}.   Thus we learn the interesting fact that $\lambda \sim v/M_{GUT}$ works well as an explanation of frozen-in DM.  The origin of the $1/M_{GUT}$ suppression could not only be a true ``Grand Unified Theory" scale in the traditional $SU(5)$ or $SO(10)$ sense, but could alternatively be the string compactification scale $M_{KK}$ (or orbifold GUT compactification scale) or the string scale $M_{str}$ depending on the precise model in which FIMPs are implemented.  Note that the choice of $M_{KK}\sim (10^{15}-10^{16})$ GeV and $M_{str} \sim10^{16}$ GeV is well motivated by modern string model building \cite{Witten:1996mz} \cite{Friedmann:2002ty} \cite{Svrcek:2006yi}, or by orbifold GUTs \cite{orbifoldGUT}.

In the following section we argue that FIMP candidates with weak-scale masses and
couplings of this order naturally appear in many beyond-the-standard-model theories, such as string theory
with low-energy supersymmetry.  Moreover in Section~\ref{signatures} we show that such FIMP
theories can be tested and explored by experiments at the LHC and by astrophysical and cosmological
observations.

\section{FIMP Candidates}\label{candidates}

We now turn to the issue of possible FIMP candidates in well-motivated theories
of beyond-the-Standard-Model physics.

\subsection{Moduli with weak scale supersymmetry}

The FIMP may be a modulus, $T$, or modulino, $\tilde{T}$, frozen-in by a variety of possible supersymmetric interactions at the weak scale.
We consider a theory with weak scale supersymmetry where the size of the supersymmetry breaking arises from the compactification of extra spatial dimensions, and depends on $T$.  We assume that the conventional moduli problem is solved, so that after reheating at the end of inflation the $T$ abundance is negligible.   The modulus $T$ has a large vev, which is not determined until supersymmetry is broken, so that $T$ is $R$ parity even and the corresponding modulino, $\tilde{T}$, is $R$ parity odd.   The modulus and modulino are expected to pick up masses of order the supersymmetry breaking scale, which we take to be of order the weak scale $v$.  It is therefore natural to imagine that the scalar $T$ can play the role of the FIMP that is produced by freeze-in and then decays to LOSP DM, while the $R$-parity odd modulino $\tilde{T}$ can either similarly be freeze-in produced and then decay to LOSP DM, or be FIMP DM itself, depending upon the hierarchy of LOSP and modulino masses.  

Most important, the interactions of $T$ and $\tilde{T}$ are naturally of the right size to lead to freeze-in production.
These interactions are obtained by expanding the supersymmetry breaking parameters about the modulus vev.  Taking these parameters to be soft scalar masses, $m^2$, the conventionally defined $A$ and $B$ parameters, the gaugino masses $m_{\tilde{g}}$ and the $\mu$ parameter, the leading (renormalisable) interactions of the modulus are 
\begin{align*}
&m^2  \left(1+\frac{T}{M} \right)\, (\phi^\dagger \phi + h^\dagger h)&  &\mu B\left(1+\frac{T}{M}  \right) \, 
h^2& &Ay  \left(1+\frac{T}{M} \right) \phi^2 h& \\
& m_{\tilde{g}}  \left(1+\frac{T}{M} \right) \, \tilde{g} \tilde{g}& &\;\mu y\left(1+\frac{T}{M} \right) \phi^2 h^*
& &\;\;\,\mu\left(1+\frac{T}{M} \right) \, \tilde{h} \tilde{h} , &
\end{align*}
where now $T$ refers to the fluctuation of the modulus about its expectation value.
Here $\phi$ is a squark or slepton, $\tilde{h}$ a Higgsino, and $h$ a Higgs boson, and $y$ is a corresponding Yukawa coupling.  The mass scale $M$ is the compactification scale, which we take to be of order the unified mass scale, $M_u$, as in many string constructions, and numerical factors of order unity are ignored. 
For the modulino the leading interactions are
\begin{equation}
\mu \, \frac{\tilde{T}}{M} \, \tilde{h} h  \, ,
\label{eq:modinoint}
\end{equation}
as well as possible terms that arise from higher-dimension $D$-terms involving the moduli-dependent susy-breaking spurion leading to
interactions with sfermion--fermion pairs
\begin{equation}
 \frac{m_{susy}}{M} \, \tilde{T} (q \tilde{q}^\dagger, l \tilde{l}^\dagger, \tilde{h} h^\dagger).
\label{eq:modinoint2}
\end{equation}
These interactions are all of the form shown in eq.~(\ref{eq:Xint}), with 
\begin{equation}
\lambda \,\sim \,  \frac{v}{M} 
\label{eq:lambda}
\end{equation}
in each case.  Thus we discover that the modulus and modulino corresponding to the size of supersymmetry breaking 
automatically lead to FIMP-like behaviour. 
This is not the case for moduli that appear in the expansion of renormalisable couplings, since then the moduli couplings are non-renormalisable and do not lead to IR-dominated freeze-in behaviour, as we discuss in Section \ref{comments}. 

There are many possible reactions for the freeze-in mechanism involving $T$ or $\tilde{T}$, and here we give a few cases for illustration.  $\tilde{T}$ FIMP dark matter could arise from the freeze-in mechanism by decays of a LOSP chargino
\begin{equation}
 \tilde{\chi}^+ \rightarrow H^+ \, \tilde{T},
\label{eq:modulinoFI1}
\end{equation}
where $H^+$ is a charged Higgs boson.  On the other hand, if the $\tilde{T}$ FIMP is heavier than the LOSP, then DM may arise by freeze-in of $\tilde{T}$ followed by the decay of $\tilde{T}$ to LOSP dark matter,
for example
\begin{equation}
 \tilde{t} \rightarrow t \, \tilde{T},  \hspace{0.5in}     \tilde{T} \rightarrow h  \tilde{\chi}^0,
\label{eq:modulinoFI2}
\end{equation}
where $t$ and $\tilde{t}$ are the top quark and top squark, and $\chi^0$ the neutralino LOSP.
Alternatively, neutralino DM could be produced by the freeze-in of a modulus FIMP which then decays, for example 
\begin{equation}
 \tilde{t}_2 \rightarrow \tilde{t}_1 T,  \hspace{0.5in}     T \rightarrow  \tilde{\chi}^0  \tilde{\chi}^0.
\label{eq:modulusFI1}
\end{equation}
%

\subsection{Dirac neutrino masses within weak scale supersymmetry}\label{diracneu}

As a second example with weak-scale supersymmetry, consider the case that 
neutrinos are Dirac, with masses generated via the interaction 
\begin{equation}
\lambda \, LNH_u,
\label{eq:Diracnu}
\end{equation}
where $L,N$ and $H_u$ are chiral superfields containing the lepton doublet, 
right-handed neutrino singlet and Higgs fields.  The coupling matrix $\lambda$ now 
determines the neutrino masses and consequently is constrained by experiment 
to have entries with maximal values that are of order $10^{-13}$, very
close to that required for a FIMP. 
The interaction leads to the freeze-in of the right-handed sneutrino, 
$\tilde{\nu}_R$, for example via $\tilde{h}l \rightarrow \tilde{\nu}_R$ and, 
as computed by Asaka et al \cite{moroi}, can lead to a successful dark matter 
abundance when $\tilde{\nu}_R$ is the LSP\footnote{For related ideas see \cite{rhsnunontherm} and for the sterile neutrino case see e.g. \cite{rhneutrino}.}.

\subsection{FIMPs from kinetic mixing}

Additional hidden $U(1)$ factors under which no light SM states are charged are a common feature of theories
beyond-the-Standard-Model.  For example, many string constructions possess multiple hidden $U(1)$ factors arising from D-branes or
from antisymmetric tensor fields with multiplicity determined by the topology of the compactification \cite{Arvanitaki:2009fg}.    The
unique renormalisable coupling of such hidden $U(1)$'s to SM states arises via the kinetic mixing term, $\epsilon_{iY} F_i^{\mu\nu} F_{Y\mu\nu}$, with hypercharge, and in field theory such kinetic mixing can be generated at the one-loop level when super-massive states charged
under both $U(1)$'s are integrated out \cite{Holdom:1985ag}, with result
\begin{equation}
\epsilon_{iY}\simeq \frac{tr( g_Y Q_Y g_i Q_i)}{12 \pi^2} \log\left(\frac{m_1}{m_2}\right). 
\label{eq:kineticmixing}
\end{equation}
For small mass splittings between the superheavy states 
$\epsilon_{iY}\sim N\alpha/(3\pi) (\Delta m/{\bar m})$
where $N$ counts the multiplicity of heavy states, $\Delta m$ is their mass splitting, and $\bar m$ their mean mass.   Although these mixings can be often harmlessly rotated away for the photons (as the string $U(1)$'s often possess no light charged states), they imply non-trivial couplings to the MSSM states of the massive hidden photini superpartners via mixing with the bino, leading to hidden photini as candidate FIMPs \cite{Ibarra:2008kn,Arvanitaki:2009hb}.

A natural possibility is that the massive states sit at the GUT, or compactification scale with splittings arising from electroweak--symmetry--breaking effects.  In this case the field theory calculation eq.~(\ref{eq:kineticmixing}) gives $\epsilon_{iY}\sim 10^{-12}-10^{-15}$, with the higher values applying if the compactification, or effective GUT scale, occurs at $10^{14}-10^{15}$~GeV as for orbifold GUTs and many semi-realistic string compactifications.  For both the heterotic string \cite{Dienes:1996zr} and type-II string theory \cite{Abel:2008ai} kinetic mixing with $U(1)_Y$ arises by a process that generalises this field theory calculation.   For example in the type-II case stretched open string states with one end on the SM brane stack and the other on the hidden D-brane lead to massive states charged under both $U(1)$'s, and a one loop open string diagram then, in general, generates kinetic mixing \cite{Abel:2008ai}.  The resulting mixing is model dependent, but is exponentially suppressed values if the compactification is warped, or if the mediating fields are massive, e.g., due to fluxes.  For either case if $\epsilon_{iY}\sim 10^{-12}-10^{-15}$ the hidden photini can be FIMPs.

\subsection{Very heavy FIMPs and extra dimensions}

Dark matter with relic abundance $Y$ and mass $m$ leads to a temperature for matter-radiation 
equality that has the parametric form $T_e \sim Ym$.   Hence, from eqns.(\ref{eq:YFO}) and (\ref{eq:YFI}) 
freeze-out and freeze-in lead to 
\begin{equation}
T_{e, FO} \, \sim \frac{m'^2}{\lambda'^2 M_{Pl}},  \hspace{1in} T_{e, FI} \, \sim \lambda^2 M_{Pl}.
\label{eq:Te}
\end{equation}
For freeze-out, as $m'$ increases $\lambda'$ must also increase to maintain the observed $T_e$. 
As is well known, unitarity prevents a WIMP being much heavier than the weak scale \cite{Griest:1989wd}.  The situation 
is completely different for freeze-in; remarkably $T_e$ is independent of the FIMP mass, which can 
therefore take any extremely large value, being bounded only by the reheat temperature after 
inflation.   Furthermore, the very feeble coupling of the FIMP is a generic feature of freeze-in
\begin{equation}
\lambda \, \sim \left( \frac{T_e}{M_{Pl}} \right)^{1/2}.
\label{eq:lambdaTe}
\end{equation}

Suppose there is some stable particle, $X$, much heavier than the weak scale.  If it has a 
coupling to SM particles, for example via a quartic scalar interaction $\lambda X^{\dagger} 
X H^{\dagger} H$, that allows it to reach thermal equilibrium then it overcloses the 
universe.  To avoid this, the coupling must be reduced; in fact, it will overclose the universe 
until the coupling is reduced far enough that it satisfies eq.~(\ref{eq:lambdaTe}) and the heavy
stable particle becomes FIMP dark matter.  How might such a small interaction arise?  If the SM
is localised on a brane in a higher dimensional manifold while the heavy particle $X$ lives 
in the bulk, then a small coupling can arise from the $X$ profile having a small value at the 
location of the SM brane.  Very small couplings occur easily since the profiles are frequently 
exponential or Gaussian.

In the multiverse there may be an environmental requirement that the DM abundance not 
be much larger than we observe.  For example, axionic DM with a very large decay constant typically
overcloses the universe, but this may be avoided by environmental selection of a small initial 
vacuum misalignment angle of the axion field \cite{axionmiss}, yielding axion DM.  Similarly, 
the bulk mass of the very heavy stable particle $X$ may be selected to give a 
sufficiently steep profile that the coupling $\lambda$ obeys eq.~(\ref{eq:lambdaTe}),
thus creating FIMP dark matter.

\section{Experimental Signatures of Freeze-In \& FIMPs}\label{signatures}

No matter what the underlying theory for freeze-in, the coupling of the FIMP
to the thermal bath is very small, so a crucial question is whether the freeze-in mechanism 
can be tested by measurements at accelerators or by cosmological observations.  
In this section we outline several possible signals although we leave the details to a companion paper \cite{companion}.
In addition to the FIMP, the freeze-in mechanism typically requires a LOSP and, since we consider 
the LOSP mass to be broadly of order the weak scale, the LOSP freeze-out process cannot 
be ignored. 
\begin{figure}[t]
\centerline{\includegraphics[width=12cm]{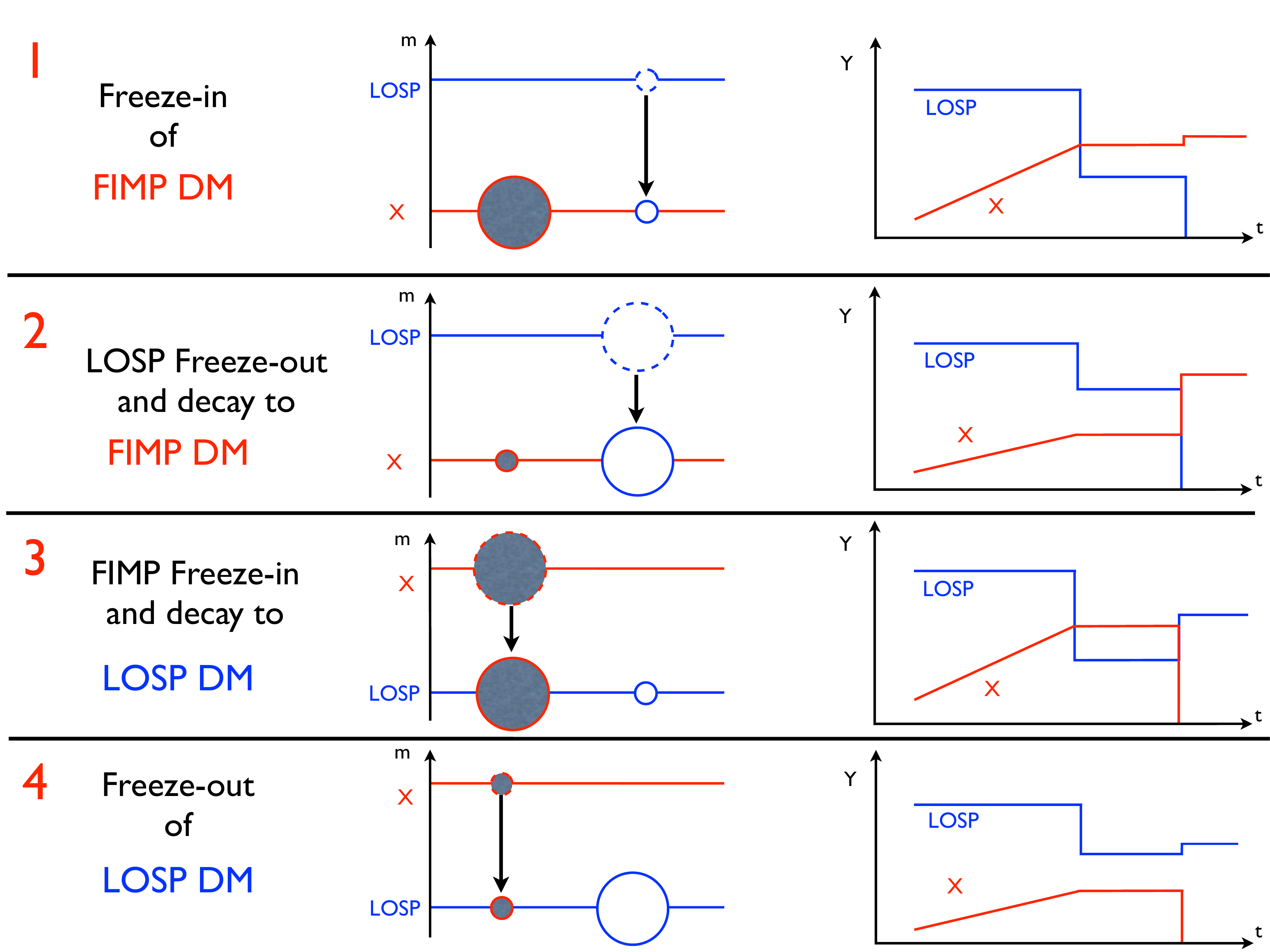}}
\caption{Schematic representation of the four possible scenarios involving the freeze-in mechanism. 
The left-hand figures show the LOSP/FIMP spectrum with circles representing cosmologically 
produced abundances. The large (small) circles represent the dominant (sub-dominant) 
mechanism for producing the dark matter relic abundance, 
a dotted (solid) circle signifies that the particle is unstable (stable), 
and a filled (open) circle corresponds to production by freeze-in (freeze-out).   The right-hand 
figures show the LOSP and FIMP abundances as a function of time.   The initial era has a thermal 
abundance of LOSPs and a growing FIMP abundance from freeze-in.  The LOSP and FIMP are 
taken to have masses of the same order, so that FIMP freeze-in ends around the same time as 
LOSP freeze-out.  Considerably later, the heavier of the LOSP and FIMP decays to the lighter.\label{scenarios}}
\end{figure}
Thus the nature of the signals depends on whether DM is dominantly produced via 
freeze-in of the FIMP or freeze-out of the LOSP and whether DM is the FIMP 
or the LOSP.   This gives four scenarios which are summarised pictorially in Figure \ref{scenarios}. 

Scenarios 1 and 2 (3 and 4) have FIMP (LOSP) dark matter, while cases 1 and 3 (2 and 4) have dark matter produced dominantly by freeze-in (freeze-out). 
Two schematic figures are shown for each of the four scenarios; the first gives the FIMP and LOSP spectrum, 
illustrating both freeze-in and freeze-out contributions to the dark matter. 
The figures on the right-hand side show the time evolution of the abundances of both the LOSP and the FIMP.  Logarithmic scales are used so that 
the freeze-out and LOSP decay processes appear very sharp, while freeze-in occurs gradually. 
The masses of the particles involved in freeze-out and freeze-in have the same order of magnitude, 
so that freeze-out and the end of freeze-in occur at comparable times.  At a later time, when 
the heavier of the LOSP and FIMP decays to the lighter, the abundance of the dark matter 
particle is boosted by the amount of the decaying particle abundance.

The four scenarios are labelled according to the dominant DM production mechanism
and the nature of DM.
\begin{enumerate}
\item {\bf Freeze-in of FIMP DM}  The FIMP is the DM and the dominant contribution to the relic DM abundance is generated via the freeze-in mechanism. 
A small abundance of LOSP freezes-out which then decays late to FIMP dark matter. 
\item {\bf LOSP freeze-out and decay to FIMP DM} The FIMP is again the DM but now the dominant contribution to the relic abundance is generated via the conventional freeze-out of the 
unstable LOSP which then decays to the FIMP.   A sub-dominant component of FIMP DM arises from freeze-in. 
\item {\bf FIMP freeze-in and decay to LOSP DM}
The LOSP is the DM and the dominant contribution to the relic abundance comes from the freeze-in of a long lived FIMP which later decays to the LOSP.  A sub-dominant component of DM arises from LOSP freeze-out.
\item {\bf Freeze-out of LOSP DM} 
The LOSP is again the DM but the dominant contribution to the relic abundance comes from conventional freeze-out of the LOSP. A small abundance of FIMPs freezes-in and decays to give a sub-dominant component of LOSP DM.  In the limit that this freeze-in abundance is small, the standard case of LOSP freeze-out is recovered. 
\end{enumerate}

The accelerator and cosmological signals depend on the scenario.  For example, production and decay of LOSPs at the LHC is possible in scenarios 1 and 2, while late decays during the MeV era of the early universe are possible in all scenarios.

\subsection{Long-lived LOSP decays at the LHC}

The hypothesis of dark matter generation via thermal
freeze-out, e.g. WIMP dark matter, may be experimentally verified by a
comparison of the observed dark 
matter density, i.e. $\Omega_{\rm DM}h^2$, and
properties of the WIMP measured at accelerators. 
The accelerator measurements
may be used to infer the WIMP self-annihilation cross 
section $\sigma_{\rm WIMP}$ which, in the simplest cases,
may be immediately employed to predict $\Omega_{\rm WIMP}h^2$. 

Does a similar correlation exist for freeze-in? Freeze-in scenarios of dark matter typically
contain both a stable dark matter particle and a very long-lived unstable state.  
In the case of FIMP dark matter, the unstable state is the LOSP.  Since 
the LOSP can only decay dominantly via the same small coupling $\lambda$ that induces freeze-in, 
then there will be a relation between the LOSP properties, in particular its mass 
and lifetime, and the FIMP dark matter abundance.    The crucial signature 
therefore arises from the accelerator production of the LOSP, followed by the 
observation of LOSP decay with a long lifetime\footnote{Investigations into the possible detection of long lived states at CMS are in progress, see for example \cite{cmslonglived}.}.  This signal is expected only for scenarios 
1 and 2;  in scenarios 3 and 4 it is the FIMP that is unstable, but the FIMP production
rate at the LHC will be negligible.  

The connection between the LOSP lifetime and the cosmological 
FIMP abundance is model dependent.  For example, suppose that the LOSP is a 
scalar $B$ and the FIMP freeze-in process occurs via a quartic scalar interaction 
leading to $BB \rightarrow XX$.  In this case the LOSP decay must involve other 
small couplings that connect $X$ to bath particles, so the relation between DM abundance
and LOSP lifetime is lost.   Here we concentrate on the 
particularly simple case that freeze-in occurs via the Yukawa coupling $\lambda B_1 B_2 X$,
where $B_1$ is the LOSP and $B_2$ a lighter bath particle.  The stability of $X$ 
can be guaranteed by a $Z_2$ parity under which $X$ and $B_1$ are odd.  
The relic abundance of FIMP dark matter arises from LOSP decays, $B_1 \rightarrow B_2 X$, 
and is computed in section 6.1 with the result of eq.~(\ref{eq:om12}).
Imposing the WMAP constraint on the dark matter abundance, eq.~(\ref{eq:om12}) can 
be rearranged to yield a prediction for the LOSP lifetime
\beq
\tau_{B_1} \, =\,7.7\times10^{-3}\mbox{sec}\, g_{B_1} \left( \frac{m_X}{100 \, \mbox{GeV}} \right) 
\, \left( \frac{300 \, \mbox{GeV}}{m_{B_1}} \right)^2\left(\frac{10^2}{g_*(m_{B_1})}\right)^{3/2}.
\label{eq:tauB1}
\eeq
This result applies for scenario 1, where the DM abundance is dominated by freeze-in.  
In scenario 2 most DM arises from LOSP freeze-out, so that the fraction from freeze-in, $f$, 
is small, $f \ll 1$.  The collider signal from late decaying LOSPs persists, but now with 
a lifetime enhanced compared with eq.~(\ref{eq:tauB1}) by a factor $1/f$. 

\subsection{Reconstructed LOSP properties at the LHC}

As illustrated in Figure \ref{scenarios}, regardless of the nature of DM (FIMP or LOSP)
the LOSP will undergo conventional freeze-out.  Consequently the dark matter abundance 
has two components, one from the freeze-in mechanism and the other from the 
freeze-out of the LOSP. The freeze-in contribution dominates only if the conventional 
thermal freeze-out cross section is large, 
$\sigma v > 3 \times 10^{-26} \mbox{cm}^3\, s^{-1} $, 
corresponding to scenarios 1 and 3.

Experiments at colliders may eventually measure the properties of the LOSP sufficiently 
well to determine that the annihilation cross section is indeed too large to 
give the observed dark matter from LOSP freeze-out, implying that the dominant contribution 
to the dark matter abundance is generated via an alternative mechanism such as freeze-in.

It would be particularly interesting if LHC discovered prompt cascade decays of superpartners
that terminated in a charged or coloured LOSP. Indeed, this would motivate a determined effort to 
search for a long lifetime associated with a LOSP, as in the previous sub-section.

\subsection{Enhanced direct and indirect detection signals of DM}

Consider the scenario where the LOSP is the dark matter and the freeze-in of a FIMP which later decays generates the correct relic abundance of LOSP dark 
matter (scenario 3 of Figure \ref{scenarios}). As discussed above this means that the conventional thermal freeze-out cross
section for the LOSP DM is large: $\sigma v > 3 \times 10^{-26} \mbox{cm}^3\, s^{-1}$.  Hence, the indirect signals produced by LOSP DM annihilating via these large cross sections can be enhanced compared to the signals predicted for DM generated directly from freeze-out.

Indirect detection experiments such as PAMELA\cite{pamela}, FERMI \cite{fermi} and
HESS \cite{hess} have recently reported deviations from background expectations of the 
proportion of positrons in cosmic rays. A common 
though perhaps premature interpretation of these experiments is the 
production of positrons resulting from WIMP annihilations. In order to match the observed fluxes one needs ``boost factors'' in the rates. Freeze-in provides a new 
avenue for building models containing boosted WIMP annihilation cross sections.

We also note that increasing the interaction strength of the dark matter particles can lead to an increase in the scattering cross section relevant in direct 
detection experiments. This could provide further evidence in favour of the freeze-in mechanism if a signal was inferred in a region of parameter space not 
consistent with frozen-out dark matter.

\subsection{Cosmological decays during the MeV era \& perturbed BBN abundances}

As has been already discussed in some detail, freeze-in scenarios usually
contain a metastable particle. This may be either the FIMP $X$ itself or the
LOSP $B_1$, 
whichever particle is the heavier one. It is well known that the
existence of metastable particles in the early Universe may be constrained
by the epoch of BBN. The light element synthesis
of $^2$H, $^3$He, $^4$He and $^7$Li (as well as $^6$Li) may change drastically
when hadronically and electro-magnetically interacting
non-thermal particles are injected into the plasma~\cite{Jedamzik:2006xz}, 
due to the decay of
the metastable particles. Significant deviations of the results of a
standard BBN scenario occur only for decay times 
$\tau\, {}^{>}_{\sim}\, 0.3\,$sec. Comparing this to eq.~(\ref{eq:tauB1}), 
and assuming
a weak mass scale FIMP such that $g_*^S\approx g_*^{\rho}\simeq 100$
as well as $g_{B_1}\approx 1$, one
arrives at a typical decay time of $\sim 10^{-2}\,$sec, implying no
signatures from BBN. However, in case the FIMP particle couples to a large
number of bath particles, i.e. $g_{B_1}\approx 100$, individual LOSP decay
times may become $\sim 1\,$sec, potentially perturbing BBN.
 
Even larger deviations (i.e. increases by factor $\sim 10^2-10^3$) from the
life time as given in eq.~(\ref{eq:tauB1}) may result if the assumed
2-body decays in eq.~(\ref{eq:tauB1}) are 
forbidden kinematically. As an example, consider the Yukawa interaction 
$\lambda X B_1 B_2$ and a mass ordering $m_{B_1} > m_X$, such that
the FIMP X is the dark matter. Assuming $m_{B_1} < m_{B_2} + m_X$
any frozen-out LOSPs $B_1$ may only decay via a 3-body decay since
$B_2$ has to remain virtual. 
Three-body decay widths are suppressed compared to 2-body decay widths resulting in longer decay times.  Given this, the typical decay time could move into the range
$\tau_{B_1} \, {}^{>}_{\sim}\, 3 - 3000\,$sec, the regime where BBN may be significantly perturbed.

\subsection{Generating a warm DM component: Erasure of small scale density perturbations}

Decay produced particle DM is often warm/hot, i.e. is endowed with primordial
free-streaming velocities leading to the early erasure of perturbations, due
to the kinetic energy imparted on the decay product during the decay 
itself.
As at least part of the dark matter in freeze-in scenarios is
produced by the late decay of the LOSP to the FIMP, or vice versa, and since
neither LOSPs nor FIMPs may thermalize below cosmic temperatures 
$T\, {}^{<}_{\sim}\, 1-10\,$MeV, freeze-in scenarios may come in the flavour
of warm or mixed (i.e cold and warm) DM scenarios. How warm depends strongly
on the decay time, and the mass ratio of mother and daughter particle. 

\section{Abundance ``Phase Diagrams"}\label{phasediagrams}

In this paper we argue that freeze-in can provide a possible alternative thermal production mechanism to the much studied freeze-out mechanism.  We now sketch ``abundance phase diagrams" that show regions of mass and coupling where each mechanism dominates the production of the relic abundance. We find that the following simple framework allows four distinct physical behaviours to account for the dominant production of dark matter.
There is a thermal bath that contains species that are kept thermally coupled during the era of interest by interactions with strength of order unity.  In addition there is a particle, $X$, that couples to the bath particles by a renormalisable interaction with a coupling $\la$.  

We assume that there is an unbroken symmetry that leads to the stability of the dark matter particle, which may be $X$ or a bath particle. This symmetry is carried by some of the bath particles, and possibly also by $X$.  As the coupling between $X$ and the bath decreases, late decays often become relevant for the production of dark matter: decay of any $X$ that is produced when $X$ is not the dark matter,  or decays to $X$ if $X$ is the dark matter.  

The interaction between $X$ and the bath has three possible forms: a quartic or trilinear scalar interaction, or a Yukawa interaction.  Furthermore the final dark matter abundance depends on whether $X$ is a fermion or boson, and on how many $X$ fields appear at the interaction.  For each such scheme one can compute the dark matter abundance in terms of the coupling and the masses of the relevant particles, and we find that there are just four possible behaviours 
\begin{itemize}
\item (I) Freeze-out of $X$.
\item (II) $X$ decouples with a full ``$T^3$" number density.
\item (III) Freeze-in of $X$.
\item (IV) Freeze-out of bath particle.
\end{itemize}
Hence, the parameter space of any such theory can be split into four ``phases" according to which behaviour yields the dominant dark matter abundance. Two of these phases correspond to the well-known case of freeze-out (I and IV). However, the introduction of very small renormalisable couplings to a theory introduces the possibility of two new phases; one with freeze-in (III) and the other having decoupling with a full relativistic abundance (II).

The above list is ordered so that, as the interaction strength between $X$ and the bath is decreased, one passes down the list, although all four ``phases" do not appear in all theories.  Below we give two illustrative examples of this ``phase" behaviour, then we argue that for very small values of $\lambda$ the ``phase" structure takes a universal form.

\subsection{Phase diagram for a quartic interaction}

\begin{figure}[tb]
\vspace{-2cm}
\centerline{\includegraphics[width=14cm]{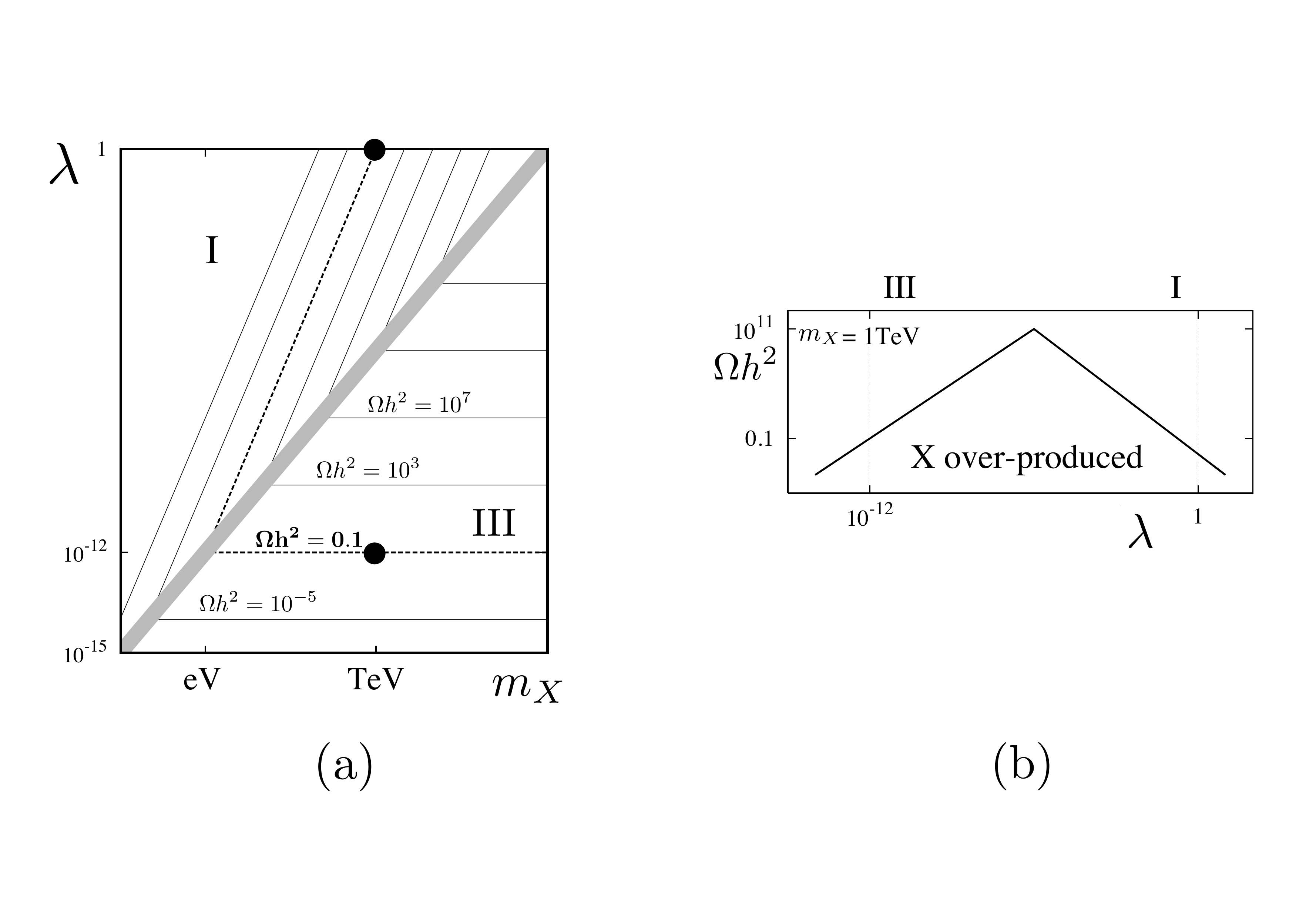}}
\vspace{-1cm}
\caption{\label{quartic} In (a) we show the contours of $\Omega h^2$ as a function of the mass $m_X$ and coupling $\la$ for the case of a quartic interaction. The plane can be divided into two ``phases": $X$ freeze-out, phase (I), occurs for large coupling and $X$ freeze-in, phase (III), occurs for weak coupling.  In (b) we take a slice at fixed $m_X \sim v$ and plot the variation of $\Om h^2$ as a function of the coupling $\la$.}
\end{figure}

Suppose that $X$ is a scalar and the interaction with the bath is via a quartic coupling to a bath scalar B
\begin{equation}
L_Q \, = \, \lambda \, B^\dagger B X^\dagger X.
\label{eq:quartic}
\end{equation}
The stabilising symmetry is a $Z_2$ with $X$ odd and $B$ even.  $X$ is the lightest particle with odd charge, and hence is stable and yields the dark matter\footnote{The phenomenology of this interaction has been previously considered in ref. \cite{mcdonald}.}. $B$ is unstable, decaying rapidly to other bath particles.  $X$ has no Standard Model gauge interactions, and has a cosmological relic abundance determined by the interaction eq.~(\ref{eq:quartic}).  

For simplicity we consider only the case $m_X > m_B$, so that the relic $X$ abundance depends on $(m_X,\lambda)$.  In Figure \ref{quartic}a contours of $\Omega h^2$ are shown in this plane.  The plane can be divided into two ``phases": $X$ freeze-out, phase (I) occurs for large coupling, $\lambda^2 > m_X/M_{Pl}$, above and to the left of the diagonal line; while $X$ freeze-in, phase (III), occurs for weak coupling with $\lambda^2 < m_X/M_{Pl}$.  These are the only two behaviours, there are no regions with phases (II) and (IV), and the plane is equally divided in logarithmic space between them.  The parametric equations for the abundances in these regions are given by eqns.~(\ref{eq:YFO}) and (\ref{eq:YFI}), with $m=m_X$.  The phase boundary occurs when $Y_{FO} \sim Y_{FI} \sim 1$: for freeze-out, $\lambda$ is sufficiently small that freeze-out occurs right at $T_{FO} \sim m_X$ and not below, while for freeze-in $\lambda$ is sufficiently large that a full thermal abundance just freezes-in by $T_{FI} \sim m_X$.  In both phases, $Y$ drops with distance from the boundary.   In the freeze-out region contours of fixed $\Omega h^2$ are straight lines:  $\lambda^2 \propto (1/\Omega h^2) (m_X/M_{Pl})^2$, while in the freeze-in region these contours are independent of $m_X$:  $\lambda^2 \propto \Omega h^2$.  Throughout the entire plane the final abundance is set in the radiation dominated era; although along the phase boundary the abundance is being set during the transition from radiation to matter domination.   

The observed value of $\Omega h^2$ is shown by a contour with a dotted line.  Along this contour, the transition from freeze-out to freeze-in occurs for a particle of mass $v^2/M_u$ just freezing-in a thermal abundance and immediately freezing-out at a temperature $v^2/M_u$,
where $M_u \sim 10^{16}$ GeV is the scale of gauge coupling unification.  This special case gives hot dark matter and is excluded; it is hard to engineer as it would require $X$ to be interacting with electrons, photons or neutrinos at the $eV$ era.  Moving away from this special point on the contour, the dark matter mass increases,  becoming warm and then cold dark matter, whether on the freeze-out or freeze-in side.

There are two particularly interesting parts of this contour: the first has $\lambda \sim 1$ and $m_X \sim v$, corresponding to the well-known WIMP case where the physics of dark matter depends on a single scale, $v$.   The second has $\lambda \sim v/M_u$ and is independent of $m_X$.  This is the freeze-in region, and within it, the case of $m_X \sim v$ is particularly interesting since in this case there are only two scales associated with the dark matter, the weak and unified scales.   These two special cases, both having $m_X \sim v$, are shown by solid dots.  One has $\lambda \sim 1$ and corresponds to the WIMP case, while the other has 
$\lambda \sim 10^{-12} \sim v/M_u$, and is the FIMP case.  A slice through the parameter space with $m_X \sim v$ is shown in Figure \ref{quartic}b, which plots the prediction for $\Omega h^2$ as a function of $\lambda^2$.   While the WIMP case is presumably related to the physics of electroweak symmetry breaking, FIMP freeze-in with $m_X\sim v$ and $\lambda \sim v/M_u$ may be related to the moduli of extra dimensional theories at the unified scale, such as string theory.  Providing the initial abundance of the moduli is suppressed, for example by well-known solutions to the moduli problem, a modulus or modulino can be frozen-in during the electroweak era to form dark matter.  For both WIMP freeze-out and moduli freeze-in, the amount of dark matter is parametrically given by
\begin{equation}
T_{eq} \, \sim \frac{v^2}{M_{Pl}}  
\label{eq:Teq}
\end{equation}
where no distinction is made here between $M_u$ and $M_{Pl}$. 

\subsection{Phase diagram for a Yukawa interaction}

\begin{figure}[t]
\vspace{-2cm}
\centerline{\includegraphics[width=14cm]{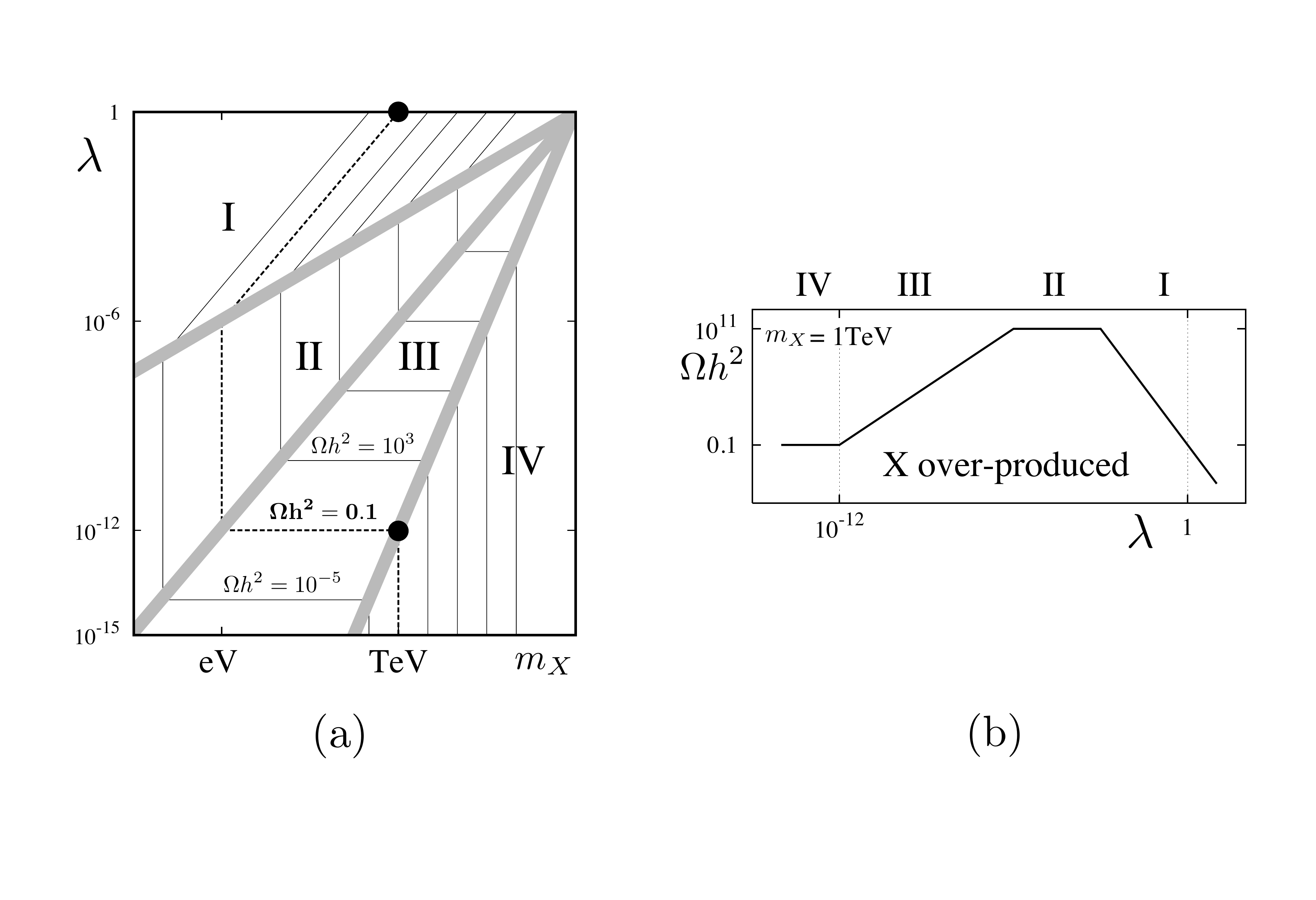}}
\vspace{-1cm}
\caption{{In (a) we show the abundance phase diagram with contours of $\Omega h^2$ as a function of the coupling $\la$ and the mass $m_X$ for the case of a Yukawa interaction, eq.~(\ref{eq:Yukawa}).  If $\lambda^2 > \sqrt{m_X/M_{Pl}}$, $X$ undergoes conventional freeze-out giving region I; while in region II, with $m_X/m_{Pl} < \lambda^2 < \sqrt{m_X/M_{Pl}}$, $X$ decouples from the bath giving a yield $Y_X \sim 1$.  In region III, with $m_X/m_{Pl} > \lambda^2 > (m_X/M_{Pl})^2$, $Y_X$ never reaches unity and freeze-in provides the dominant contribution to dark matter. For $\lambda^2 < (m_X/M_{Pl})^2$ the dominant mechanism generating dark matter arises from $\psi_1$ freezing out and then decaying to $X +\psi_2$, giving region IV.  In (b) we take a slice at $m_X=1\, {\rm TeV}$ and plot the variation of $\Om h^2$ as a function of the coupling $\la$.\label{yukcom}}}
\end{figure}

Suppose that the scalar, $X$, interacts with the bath via a Yukawa interaction with two fermions 
\begin{equation}
L_Y \, = \, \lambda \, \psi_1 \psi_2 X,
\label{eq:Yukawa}
\end{equation}
where $m_2 \ll m_X < m_1$, with $m_X$ and $m_1$ of the same order of magnitude. The stabilising symmetry is a $Z_2$ carried by both $X$ and $\psi_1$, which is the lightest bath particle odd under the symmetry.  Hence, the dark matter is $X$ and it has an abundance with an order of magnitude determined by $(m_X,\lambda)$.   Contours of $\Omega h^2$ are shown in Figure \ref{yukcom}a, and all four phases discussed above occur, in the regions labelled (I), (II), (III) and (IV).

The initial $X$ abundance is assumed to be negligible, and as we progress 
from phase I to phase IV the freeze-in process becomes successively less 
efficient.  If $\lambda^2 > m_X/M_{Pl}$, corresponding to regions I and II, 
a full thermal $X$ abundance is produced at temperatures above $m_X$.  
If $\lambda^2 > \sqrt{m_X/M_{Pl}}$, $X$ undergoes conventional freeze-out via 
the reaction $XX \rightarrow \psi_2 \psi_2$, giving region I; while in 
region II, with $m_X/M_{Pl} < \lambda^2 < \sqrt{m_X/M_{Pl}}$, $X$ decouples 
from the bath at $T>m_X$, giving a yield $Y_X \sim 1$.   On the other hand, 
with $\lambda^2 < m_X/M_{Pl}$, the freeze-in process, 
$\psi_1 \rightarrow \psi_2 X$, is less efficient, so that $Y_X$ never 
reaches unity in regions III and IV.  In region III, with 
$m_X/M_{Pl} > \lambda^2 > (m_X/M_{Pl})^2$, freeze-in provides the dominant 
contribution to dark matter, while, for 
$\lambda^2 < (m_X/M_{Pl})^2$ the dominant contribution to dark matter arises 
from $\psi_1$ freeze-out, $\psi_1 \psi_1 \rightarrow$ lighter bath particles, 
followed by $\psi_1 \rightarrow X \psi_2$, giving region IV.

Figure \ref{yukcom}b shows a slice through the two dimensional parameter space
of Figure \ref{yukcom}a with $m_X$ fixed to 100 GeV.  For a large range of $\lambda$ too much $X$ 
dark matter is produced.  But, as with the quartic interaction, there are two very interesting values of the 
coupling which yield the observed abundance of dark matter, WIMPs with $\lambda \sim 1$ and 
FIMPs with $\lambda \sim 10^{-12}$.  For these two 
special cases $\Omega h^2$ passes through the observed value, as shown in 
by solid dots Figure \ref{yukcom}a.  FIMPs with a weak-scale mass lie close to the phase boundary 
between $X$ freeze-in and $\psi_1$ freeze-out.   This important case, 
motivated, for example, by the possibility that $X$ is a modulus, will 
be examined in some detail in the next two sections.  Many of the 
observational consequences of weak-scale FIMPs arise because there are two broadly 
comparable production mechanisms, involving both freeze-in and late decays.

\begin{figure}[t]
\vspace{-0.5cm}
\centerline{\includegraphics[width=9cm]{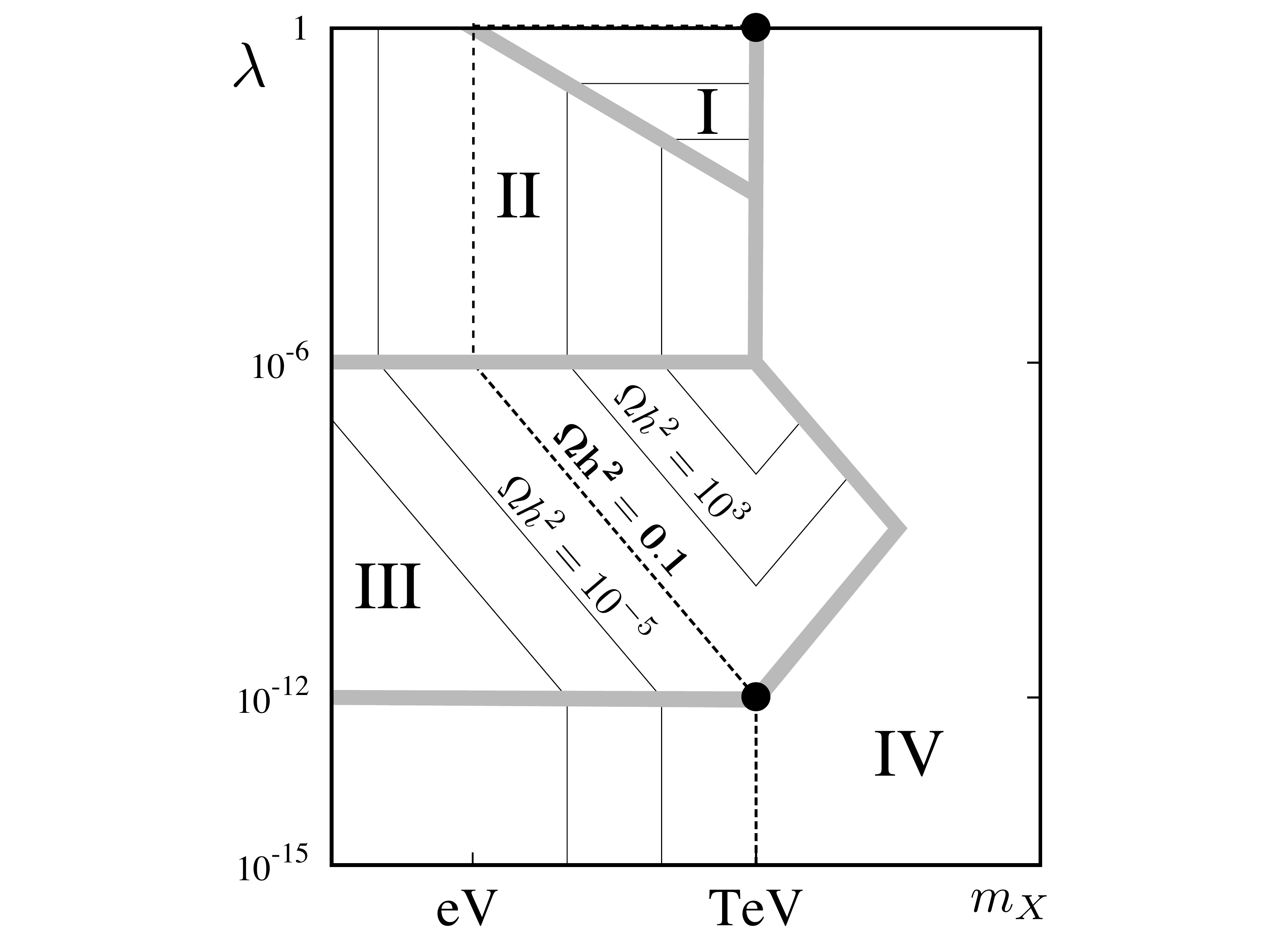}}
\caption{Fixing the bath particle mass $m_1=1\, {\rm TeV}$ the phase 
diagram shows more structure. 
For $m_X<m_1$, $X$ is 
the dark matter as in Figure \ref{yukcom}.
A new feature compared to Figure \ref{yukcom} emerges 
for $m_X>m_1$, where $\psi_1$ is the dark matter. In this case dark matter
can be dominantly produced by the freeze-in of $X$ at 
temperatures $T > 1\,$TeV, and the subsequent decay of 
$X\to \psi_1+\psi_2$. Here the decay has to occur after the freeze-out of
$\psi_1$, otherwise the $\psi_1$ contribution due to $X$-decay will be
reprocessed. The $X$-decay contribution of $\psi_1$ dark matter dominates
the $\psi_1$ freeze-out contribution in the triangle defined by the boundaries
of zone III and the condition $m_X > m_1$. A slice through this figure with $m_X$
fixed at a TeV gives the same dependence as shown in Figure \ref{yukcom}b.\label{yukmassfix}}
\end{figure}

In Figure \ref{yukcom}, we have $m_X$ and $m_1$ of the same order of magnitude as we scan over different values for $m_X$. We can develop this scenario by fixing  $m_1=1\, {\rm TeV}$ while still allowing $m_X$ to vary. The resulting phase diagram is shown in Figure \ref{yukmassfix}. Once again the diagram is split into four different regions but now these regions, in particular region IV, have more structure. The first feature to notice is the vertical dividing line at
$m_X=m_{1}=1\,{\rm TeV}$ which separates the two cases where $m_X$ is the lightest and therefore the dark matter (left hand side) and $m_{1}$ is the lightest and therefore the dark matter (right hand side). 

As before, region I corresponds to the conventional freeze-out of $X$ via the 
reaction $XX \rightarrow \psi_2 \psi_2$ and region II corresponds to where 
the $X$ particle decouples from the thermal bath with a yield $Y_X \sim 1$.  
Both these regions are cut off at $m_X=m_1$ due to the fact that when 
$m_X>m_1$, $X$ is no longer stable and decays to bath particles. Consequently, the dark matter abundance arises from $\psi_1$ freeze-out, $\psi_1 \psi_1 \rightarrow$ lighter bath particles.  

Moving down the diagram to smaller values of the coupling $\la$, 
region III is reached and the freeze-in mechanism dominates. 
For $m_X<m_1$, $X$ is the dark matter with the freeze-in process, 
$\psi_1 \rightarrow \psi_2 X$, generating the dominant contribution to 
the relic abundance.  Moving into the region where $m_X>m_1$ we now freeze-in 
an abundance of $X$ particles via the process $\psi_1+\psi_2 \rightarrow X$. 
The resulting abundance of $X$ particles then decays back to $\psi_1$ which 
forms the dark matter. Within this region as we move to larger values of 
$m_X$ at constant $\la$ we decrease the abundance of 
$X$ particles frozen-in but we also decrease the lifetime of the $X$ 
particles. Eventually, the lifetime 
becomes so short that the $X$ particles will decay back to the bath particles 
before $\psi_1$ freezes out. This means that the relic abundance of $\psi_1$ 
particles is once again determined by the freeze-out of $\psi_1$.

Moving to even smaller $\la$ we move into region IV where the freeze-in 
mechanism becomes less efficient. For $m_X<m_1$ the relic abundance of $X$ 
particles is determined by the freeze-out density of $\psi_1$ particles which 
then decay to $X$ particles after freezing out. For $m_X>m_1$ the $X$ 
particles play no role in determining the relic abundance of dark matter. 
The $\psi_1$ particle is the dark matter with it's relic abundance determined 
by conventional freeze-out.

\subsection{A universal phase diagram at small coupling}
\begin{figure}
\vspace{-1cm}
\centerline{\includegraphics[width=12cm]{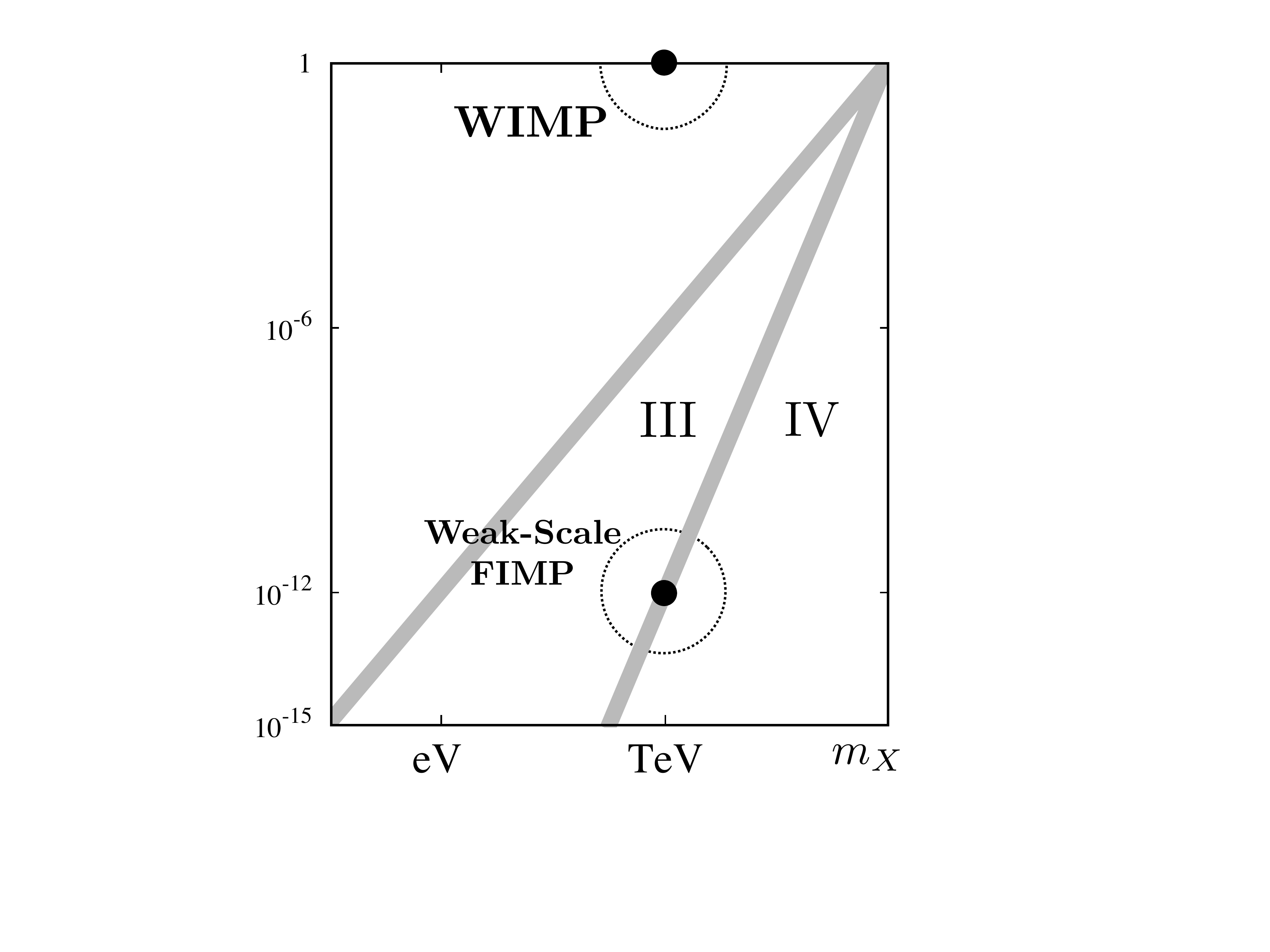}}
\vspace{-1cm}
\caption{Independent of the interaction, there is universal behaviour below the upper shadowed line, corresponding to small coupling, $\lambda^2 \ll m_X/M_{Pl}$.  For such couplings the dominant production of dark matter arises from freeze-in of $X$ if $\lambda > m_X/M_{Pl}$, and freeze-out of a bath particle if $\lambda < m_X/M_{Pl}$.\label{universal}}
\end{figure}
In Figure \ref{yukcom}a the two phases at lowest coupling correspond to $X$ freeze-in, III, and $\psi_1$ freeze-out, IV.  We find that this phase structure at small coupling results in a very wide class of theories.  Consider theories where $X$ is either a scalar, $A_X$, or a fermion, $\psi_X$, with linear couplings to bath scalars (A) and fermions ($\psi$) that are renormalisable\footnote{For simplicity we ignore the quadratic interactions $\lambda A_X \psi_X \psi$,  $\lambda A_X^2 A^2$, $\lambda A \psi_X \psi_X$, and $\lambda m_X A_X^2 A$.  A similar analysis holds in the more general case.}, so that the possible interactions are
\begin{equation}
\lambda A_X \psi \psi, \qquad \lambda m_X  A_X A^2, \qquad \lambda A_X A^3, \qquad {\rm and} \qquad \lambda \psi_X \psi A .
\label{eq:Xint}
\end{equation}

The stabilizing symmetry is $Z_2$, and $A_X, \psi_X$ may be even or odd (although, if it is even, it must have a significant fraction of its coupling to bath states that are odd).   Any choice of parity may be made for $A$ and $\psi$ as long as the interactions of eq.~(\ref{eq:Xint}) are symmetric.  We assume that $Z_2$-odd bath particles in the interactions of eq.~(\ref{eq:Xint}), as well as the lightest bath particle odd under the $Z_2$, $B$, have masses of the same order of magnitude as $m_X$, and that $Z_2$-even bath particles in the interactions of eq.~(\ref{eq:Xint}) do not  have masses parametrically larger than $m_X$.   

For this wide class of theories, the phase diagram is shown in Figure \ref{universal}, for $\lambda^2 \ll m_X/M_{Pl}$.  The dominant production of dark matter, whether $X$ or bath,  arises from freeze-in of $X$, if $\lambda > m_X/M_{Pl}$, and freeze-out of a bath particle, if $\lambda < m_X/M_{Pl}$.  This generality follows from two results.  Since $B$, the lightest $Z_2$-odd bath particle, has order unity couplings to lighter bath particles, its freeze-out yield is always $Y_B \sim m_B/M_{Pl} \sim m_X/M_{Pl}$.  Freeze-in of $X$ via $2\rightarrow 2$, $1 \rightarrow 2$ or $ 2 \rightarrow 1$,  involving any quartic, trilinear or Yukawa interaction of eq.~(\ref{eq:Xint}), leads to a yield $Y_X \sim \lambda^2 M_{Pl}/m_X$.

>From Figure \ref{universal} we see that this class of theories contains the interesting weak-scale FIMP case, $m_X \sim v$ and $\lambda \sim v/M_u$. This leads to a
generically realistic abundance of dark matter, arising from two broadly comparable
mechanisms---$X$ freeze-in and $B$ freeze-out---as shown by the FIMP label on the phase boundary in Figure  \ref{universal}.

\section{Freeze-In Calculation}\label{calculation}

We now turn to the calculation of the frozen-in dark matter density.   
The freeze-in process is dominated by decays or inverse decays of bath particles to the FIMP depending
on whether or not the FIMP is the lightest particle carrying the conserved quantum number that stabilises
the DM.  Processes such as $2\to2$ and other scatterings give sub-dominant contributions to the frozen-in
abundance in almost all cases of interest as the diagrams are suppressed by both additional SM gauge or
Yukawa couplings and numerical factors of order ${\rm few}\times \pi^3$ arising from the additional
phase space integrals.  


\subsection{Direct freeze-in of dark matter}

Consider the case where the FIMP $X$ is the dark matter particle itself.
If $X$ has a coupling to two bath particles, $\lambda X B_1 B_2$, then, 
for $m_{B_1} > m_{B_2} + m_X$, the dominant freeze-in process is via decays 
of the heavier bath particle:  
\begin{equation}
B_1 \rightarrow B_2 X.
\label{eq:B1decay}
\end{equation}
At temperatures much higher than $m_{B_1}$, the yield during a Hubble 
doubling time at the era of temperature T scales as 
\begin{equation}
Y_{1 \rightarrow 2}(T) \propto \frac{M_{Pl} \, m_{B_1}\Ga_{B_1}}{T^3}, 
\label{eq:YT12}
\end{equation}
so that it is strongly IR dominated, and shuts off only once the temperature 
drops below $M_{B_1}$. 

We can be more precise by solving the Boltzmann equation for $n_X$, the number 
density of $X$ particles in this case:
%
\begin{multline}
\dot{n}_X+3 H n_X =
\int d\Pi_Xd\Pi_{B_1}d\Pi_{B_2}(2\pi)^4\de^4(p_X+p_{B_2}-p_{B_1})\\
\times\left[\abs{M}^2_{B_1\rightarrow B_2+X}f_{B_1}(1\pm f_{B_2})(1\pm f_X)-
\abs{M}^2_{B_2+X\rightarrow B_1}f_{B_2}f_X(1\pm f_{B_1})\right],
\label{eq:1to2}
\end{multline}
where $d\Pi_i=d^3p_i/(2\pi)^32E_i$ are phase space elements, and $f_i$ is phase
space density of particle $i$, such that
\begin{equation}
n_i = \frac{g_i}{2\pi^3}\int{\rm d}^3p f_i
\end{equation} 
is the total particle density of species $i$ possessing $g_i$ internal spin
degrees of freedom. It is implicitly assumed
that in eq.~(\ref{eq:1to2}), as well as
eqns.~(\ref{Boltzman1}) and (\ref{eq:2to1}) below, the squares of the matrix 
elements are summed over final and
initial spin of the participating particles without averaging over the initial
spin degrees of freedom. 
We assume the initial $X$ abundance is negligible so that we 
may set $f_X=0$, such that we may neglect the second term 
in eq.~(\ref{eq:1to2}).
Using the definition for the partial decay width $\Gamma_{B_1}$ of 
$B_1\rightarrow B_2 X$ and neglecting Pauli-blocking/stimulated emission
effects, i.e. approximating $(1\pm f_{B_2})\approx 1$, 
we can rewrite the Boltzmann equation as 
\beq
\dot{n}_X+3H n_X \approx 2 g_{B_1}\int d\Pi_{B_1}\Ga_{B_1}m_{B_1}f_{B_1}=
g_{B_1}\int \frac{d^3p_{B_1}}{(2\pi)^3}\frac{f_{B_1}\Ga_{B_1}}{\ga_{B_1}}
\eeq
where $\ga_{B_1}=E_{B_1}/m_{B_1}$.
The bath particles are assumed to be in thermal equilibrium and 
so approximating 
$f_{B_1}= (e^{E_{B_1}/T}\pm 1)^{-1}$ by $e^{-E_{B_1}/T}$ 
and converting the integral over momentum space into an integral over 
energy we have
\begin{multline}
\dot{n}_X+3n_XH\approx g_{B_1}\int \frac{d^3p_{B_1}}{(2\pi)^3}\frac{f_{B_1}\Ga_{B_1}}{\ga_{B_1}}
=g_{B_1}\int_{m_{B_1}}^{\infty}\frac{m_{B_1}\Ga_{B_1}}{2\pi^2}(E_{B_1}^2-m_{B_1}^2)^{1/2}e^{-E_{B_1}/T}dE_{B_1}\\
=\frac{g_{B_1}m_{B_1}^2\Ga_{B_1}}{2\pi^2}TK_1(m_{B_1}/T).
\end{multline}
where $K_1$ is the first modified Bessel Function of the 2nd kind.
Rewriting in terms of the yield, $Y\equiv n/S$ and using 
$\dot{T}\approx -HT$, applicable when the variation of total plasma
statistical degrees of freedom with temperature 
${\rm d}g/{\rm d}T\approx 0$ approximately vanishes, we have 
\beq
Y_X\approx \int_{T_{min}}^{T_{max}} \frac{g_{B_1}m_{B_1}^2\Ga_{B_1}}{2\pi^2}\frac{K_1(m_{B_1}/T)}{SH}dT,
\eeq
with $S=2\pi^2g_*^ST^3/45$ and $H=1.66\sqrt{g_*^{\rho}}T^2/M_{Pl}$. Using $x\equiv m/T$ we can rewrite the integral as
\beq
Y_X\approx \frac{45}{(1.66) 4\pi^4} 
\frac{g_{B_1}M_{Pl}\Ga_{B_1}}{m_{B_1}^2g_*^s\sqrt{g_*^{\rho}}}\int_{x_{min}}^{x_{max}}K_1(x)x^3dx\,,
\eeq
where $M_{Pl}$ is the (non-reduced) Planck mass.
Doing the $x$ integral with $x_{max}=\infty$ and $x_{min}=0$ we finally arrive 
at the result
\begin{equation}
Y_{1 \rightarrow 2}\approx \frac{135\, g_{B_1}}{8\pi^3(1.66) g^S_* \sqrt{g^{\rho}_*}} \left( \frac{M_{Pl} \Gamma_{B_1}}{m_{B_1}^2} \right), 
\label{eq:Y12}
\end{equation}
where $g_*^{S,\rho}$ are the effective numbers of degrees of freedom  
in the bath at the freeze-in temperature $T\sim m_{B_1}$ for the entropy, 
$S$, and energy density, $\rho$, respectively. Thus the abundance depends on the three quantities $m_X, m_{B_1}$ and $\Gamma_{B_1}$ and has the form
\beq
\Om_X h^2\approx \frac{1.09\times 10^{27} g_{B_1}}{g^S_* \sqrt{g^{\rho}_*}} \frac{m_X \Ga_{B_1}}{m_{B_1}^2} .
\label{eq:om12}
\eeq
This is the density of $X$ produced by a single 
bath particle species.  In full FIMP models it is quite likely that a number of bath particles could have similar interactions with the FIMP particle.  For example, if $X$ arises as an $R_p$-odd modulino state from the SUSY-breaking sector of a string theory, then as discussed in
Section \ref{candidates}, $X$ is likely to couple to most if not all of the MSSM spectrum.   In this case 
further contributions to the relic abundance are generated with the same dependence on the bath particle's
partial decay width and mass, giving
\beq
\Om_X h^2 |_{tot}\approx \frac{1.09\times 10^{27}}
{g^S_* \sqrt{g^{\rho}_*}} m_X  
\sum_i \frac{g_{B_i}\Ga_{B_i}}{m_{B_i}^2} ,
\label{eq:om12a}
\eeq
and the overall interaction strength will need to be smaller in order to generate the correct abundance of FIMP dark matter.   If we approximate this enhancement by an effective number of active degrees of freedom, $g_{bath}$, each with mass $m_B$, then the required interaction strength becomes 
\beq
\lambda \simeq 1.5\times 10^{-13} \left(\frac{m_{B}}{m_X}\right)^{1/2}  \left(\frac{g_*(m_{B})}{10^2}\right)^{3/4}  
\left(\frac{g_{bath}}{10^2}\right)^{-1/2} .
\label{eq:la12b}
\eeq

\subsection{Decays of frozen-in FIMPs to dark matter}

The alternative possibility is that the particle that freezes-in 
is unstable and decays to dark matter.  Here we study the simplest possibility 
that the interaction responsible for freeze-in also yields the decay.   
If $X$ has a coupling to two bath particles, $\lambda X B_1 B_2$, then, 
for $m_X > m_{B_1} + m_{B_2}$, the dominant freeze-in process is via inverse 
decays of $X$:  
\begin{equation}
B_1  B_2 \rightarrow X.
\label{eq:Xinvdecay}
\end{equation}
At temperatures much higher than $m_X$, the yield during a Hubble doubling 
time at the era of temperature T scales as 
\begin{equation}
Y_{2 \rightarrow 1}(T) \propto \frac{M_{Pl} \, m_X\Ga_X}{T^3}, 
\label{eq:YT21T}
\end{equation}
and is again strongly IR dominated, shutting off as the temperature falls 
below $m_X$. 

Assuming once again that the initial abundance of $X$ particles is zero and therefore setting $f_X=0$ the Boltzmann equation for this process can be written as
\beq
\dot{n}_X+3Hn_X\approx\int d\Pi_Xd\Pi_{B_1}d\Pi_{B_2}(2\pi)^4\de^4(p_X-p_{B_1}-p_{B_2})\abs{M}^2_{B_1+B_2\rightarrow X}f_{B_1}f_{B_2}.
\label{Boltzman1}
\eeq
Assuming CP invariance we may set  $\abs{M}^2_{B_1+B_2\rightarrow X}=\abs{M}^2_{X\rightarrow B_1+B_2}$ and by invoking the principle of detailed balance we can rewrite the Boltzmann equation as
\beq
\dot{n}_X+3n_XH\approx\int d\Pi_Xd\Pi_{B_1}d\Pi_{B_2}(2\pi)^4\de^4(p_X-p_{B_1}-p_{B_2})\abs{M}^2_{X\rightarrow B_1+B_2}f^{eq}_{X},
\label{eq:2to1}
\eeq
where $f^{eq}_{X}$ is the $X$ equilibrium phase space distribution
approximated again by $f^{eq}_{X}\approx e^{-E_X/T}$. 
Comparing eqns.~(\ref{eq:2to1}) with (\ref{eq:1to2}) we can immediately 
write down the resulting form of the $X$ yield as
\beq
Y_{2 \rightarrow 1} \approx  \frac{135}{8\pi^3(1.66)  g^S_* \sqrt{g^{\rho}_*}} \left( \frac{M_{Pl} \Gamma_{X}}{m_{X}^2} \right), 
\label{eq:Y21}
\eeq
where $\Gamma_{X}$ is the partial width of $X\rightarrow B_1B_2 $. 

Assuming that $B_1$ is the DM particle, and in addition that the freeze-in contribution coming from decays of $X$ dominates the conventional freeze-out abundance of $B_1$, the final DM density is
\beq
\Om_{B_1} h^2\approx\frac{1.09\times 10^{27}}{g^S_* \sqrt{g^{\rho}_*}} \frac{m_{B_1} \Ga_{X}}{m_{X}^2} .
\label{eq:om21}
\eeq
Here, crucially, we have assumed the decay of $X\to B_1 B_2$ occurs at a time after the freeze-out of $B_1$ so that the density does not get reprocessed.
Taking $\Ga_{X}=\lambda^2 m_{X}/8\pi$, the required dark matter
density occurs for a coupling of size
\beq
\lambda \simeq 1.5\times 10^{-12} \left(\frac{m_{X}}{m_{B_1}}\right)^{1/2}  \left(\frac{g_*(m_{X})}{10^2}\right)^{3/4} .
\label{eq:la21}
\eeq
Although eq.~(\ref{eq:om21}) is very similar in form to eq.~(\ref{eq:om12}), as was to be expected, the physics is quite different.   

\subsection{Freeze-in by 2$\rightarrow$2 scattering}

Finally we present the calculation of the FIMP relic abundance in the case where the FIMP, $X$, is a scalar and interacts with three scalar bath particles $B_1$, $B_2$ and $B_3$ via the operator
\beq
\mathcal{L}_{4-scalar}=\la XB_1B_2B_3.
\eeq
Considering this interaction we can calculate the resulting FIMP yield using the Boltzmann equation
\beq
\dot{n}_{X}+3n_{X}H\approx 3\int d\Pi_{B_1}d\Pi_{B_2}d\Pi_{B_3}d\Pi_{X}(2\pi)^4\de^4(p_{B_1}+p_{B_2}-p_{B_3}-p_{X})\abs{M}^2_{B_1B_2\rightarrow B_3X}f_{B_1}f_{B_2},
\eeq
where the factor of 3 accounts for the fact that we can have $B_1B_2\rightarrow B_3X$, $B_1B_3\rightarrow B_2X$ and $B_2B_3\rightarrow B_1X$ contributing to the FIMP yield all with the same rate.
We assume that the masses of $B_1, B_2$ and $B_3$ are negligible compared to the FIMP particle mass. We can rewrite the Boltzmann equation as a one dimensional integral, \cite{edsgon},
\beq
\dot{n}_{X}+3n_{X}H\approx\frac{3T}{512\pi^6}\int_{m_X^2}^{\infty} ds \, d\Om P_{B_1B_2}P_{B_3X}\abs{M}^2_{B_1B_2\rightarrow B_3X}K_1(\sqrt{s}/T)/ \sqrt{s},
\eeq
where $s$ is the centre of mass energy of the interaction at a temperature $T$ and 
\beq
P_{ij}\equiv\frac{[s-(m_i+m_j)^2]^{1/2}[s-(m_i-m_j)^2]^{1/2}}{2\sqrt{s}}.
\eeq
The matrix element is $\abs{M}^2_{B_1B_2\rightarrow B_3X}=\la^2$ leaving
\beq
\dot{n}_{X}+3n_{X}H\approx\frac{3T \la^2}{512\pi^5}\int_{m_X^2}^{\infty} ds \, (s-m_X^2) K_1(\sqrt{s}/T)/\sqrt{s},
\eeq
Doing this $s$ integral and using the definition for the yield, $Y=n/S$ we have
\beq
\frac{dY_{X}}{dT}\approx -\frac{3\la^2T^2m_X}{SH}\frac{K_1(m_X/T)}{128\pi^5}=\frac{3\la^2K_1(m_X/T)}{1.66\,T^3g^S_* \sqrt{g^{\rho}_*}}\frac{45M_{pl}m_X}{256\pi^7}.
\eeq
Changing variables from $T$ to $x\equiv m_X/T$ and again doing the integral $x$ integral under the approximations that $x_{max}=\infty$ and $x_{min}=0$ we finally arrive at the result
\beq
Y_{X}\approx \frac{135\la^2M_{pl}}{256\pi^7g^S_* \sqrt{g^{\rho}_*}(1.66)m_X}\int_0^{\infty}xK_1(x)dx=\frac{135M_{pl}\la^2}{512\pi^6g^S_* \sqrt{g^{\rho}_*}(1.66)m_X}.
\eeq
The relic density of $X$ FIMPs is then given by
\beq
\Om h_X^2\approx\frac{2m_XY_X}{{3.6\times 10^{-9}\,{\rm GeV}}}=\frac{1.01\times 10^{24} }{g^S_* \sqrt{g^{\rho}_*}}  \la^2.
\eeq
Finally to generate the required relic abundance we need
\beq
\la \simeq 1\times 10^{-11}\left(\frac{g_*(m_{X})}{10^2}\right)^{3/4} ,
\eeq
larger than the corresponding value for the three body interactions.

\section{Comments and Discussion}\label{comments}

Having outlined the elementary theory of the freeze-in mechanism
and FIMP phenomenology we now address some of the complications
and issues that can arise in complete theories or ones where the FIMP
sector is not just one particle.

\subsection{Higher dimension operators and FIMPs from GUTs}

In addition to the example candidates presented in section \ref{candidates} an important application of the freeze-in mechanism is to non-renormalisable higher dimensional operators (HDO)s containing at least one Higgs-like state. Consider the operator
\beq
\mathcal{L}_{HDO}=\frac{\al}{M_{}^{n}}(\varphi_1\varphi_{2}...\varphi_n)X\psi_1\psi_2,
\eeq
where $\al$ is an $\mathcal{O}(1)$ coupling, $X$ is our FIMP state, $\psi_1$ and $\psi_2$ are two fermionic bath states and the $\varphi_i$s are Higgs-like states, also assumed to be in thermal equilibrium, that will gain vacuum expectation values (VEVs) at some energy scale below the high scale $M_{}$. These higher dimension $1/M$ suppressed operators involving Higgs VEVs (e.g. for $n=1$ electroweak scale VEVs or for $n>1$ larger-scale VEVs\footnote{For $n=0$, the resulting operator is renormalisable and has an $\mathcal{O}(1)$ coupling. This interaction will not lead to freeze-in due to the large coupling and so we assume it is absent due to symmetries.}) can lead to interactions similar to the dimension-4 Yukawa interaction of eq.~(\ref{eq:Yukawa}) with the appropriate coupling size to generate the correct freeze-in abundance of dark matter.

For example, dimension five $1/M_{GUT}$-suppressed operators involving the SM Higgs VEV can lead to
interactions with SM-singlets contained in GUT representations giving FIMPs with interaction strength,
$\lambda \sim v/M_{GUT} \sim 10^{-13}$.  Alternatively, dimension six $1/M_{GUT}^2$-suppressed operators involving
two intermediate scale VEV's can also give rise to suitably sized FIMP couplings\footnote{For existing models employing $1/M_{GUT}$ operators in the context of decaying dark matter see for example  \cite{mina}.}.  However if the excitations of the Higgs-like states are not super-massive and are in thermal equilibrium there is a UV contribution to the dark matter yield that limits the applicability of the IR freeze-in mechanism.

For the sake of clarity we specialise to the case of $n=1$ with one Higgs state gaining an electroweak scale VEV and mass.  Expanding $\varphi_1$ around this VEV we have the following operators
\beq
\mathcal{L}_{HDO}=\frac{\al v}{M_{}}X\psi_1\psi_2+\frac{\al}{M_{}}\varphi_1X\psi_1\psi_2.\label{hdo}
\eeq
Under the assumption that $m_{\psi_1}>m_{\psi_2}+m_{X}$ the IR dominated freeze-in yield is determined by the decay $\psi_1\rightarrow \psi_2 X$ with rate determined by the coupling $\al^{\prime}\equiv \al v/M_{}$.
The resulting yield can be directly read from eq.~(\ref{eq:Y12}) and has the form
\begin{equation}
Y \approx \frac{10.2}{\pi^3 g^S_* \sqrt{g^{\rho}_*}} 
\left( \frac{M_{Pl} \Gamma_{\psi_1}}{m_{\psi_1}^2} \right)\sim\frac{1.3 }{\pi^4 g^S_* \sqrt{g^{\rho}_*}} \left(\frac{M_{Pl} \al^{\prime^2}}{m_{\psi_1}} \right),
\end{equation}
where we have approximated the rate as $\Ga_{\psi_1}\approx 
\al^{\prime^2} m_{\psi_1}/8\pi$.  This contribution is IR 
dominated by temperatures close to the mass $m_{\psi_1}$.  However, we also get a contribution to the yield of $X$ coming from the non-renormalisable interaction in eq.~(\ref{hdo}) via a four particle interaction such as $\psi_1\varphi_1\rightarrow  X\psi_2$ and the important point is that this contribution is UV dominated, so this contribution will depend on unknown UV physics such as the reheat temperature $T_R$.  

What constraint is there on $T_R$ for the yield to be determined primarily by IR physics?
Considering the HDO interaction only we can calculate the resulting yield using the Boltzmann equation
\beq
\dot{n}_{X}+3n_{X}H\approx \int d\Pi_{\varphi}d\Pi_{\psi_1}d\Pi_{\psi_2}d\Pi_{X}(2\pi)^4\de^2(p_{\psi_1}+p_{\varphi}-p_{\psi_2}-p_{X})\abs{M}^2_{\psi_1\varphi\rightarrow \psi_2X}f_{\varphi}f_{\psi_1}.
\eeq
Manipulating this equation we can write
\beq
\dot{n}_{X}+3n_{X}H\approx\frac{T}{2048\pi^6}\int ds \, d\Om \sqrt{s}\abs{M}^2_{\psi_1\varphi\rightarrow \psi_2X}K_1(\sqrt{s}/T),
\eeq
where $s$ is the centre of mass energy of the interaction at a temperature $T$ and we have approximated the masses of the relevant particles to be negligible compared to the temperature at which we are working. In this limit the matrix element is $\abs{M}^2_{\psi_1\varphi\rightarrow \psi_2X}=\al^2/M^2s$ leaving
\beq
\dot{n}_{X}+3n_{X}H\approx\frac{T\al^2}{512\pi^5M^2}\int_0^{\infty} ds \, s^{3/2} K_1(\sqrt{s}/T),
\eeq
Doing this final integral and using the definition for the yield, $Y=n/S$ we have
\beq
\frac{dY_{UV}}{dT}\approx -\frac{1}{SHT}\frac{T^6\al^2}{16\pi^5M^2}.
\eeq
Now performing the final T integral which is dominated at the highest temperature, $T_R$.
\beq
Y_{UV}\approx \frac{0.4 \,T_R\al^2M_{pl}}{\pi^7M^2 g_{*}^S\sqrt{g_{*}^{\rho}}}.
\eeq

Thus in order for the IR contribution to dominate over that arising from the UV we need
\beq
\frac{Y}{Y_{UV}}\simeq\frac{3\pi^3 v^2}{m_{\psi_1}T_R}>1,
\eeq
translating into an upper bound on the reheat temperature given by
\beq
T_R\, _{\sim}^<\,\frac{3\pi^3v^2}{m_{\psi_1}}.
\eeq
With $m_{\psi_1}=150$GeV, the maximum value for $T_R$ is around 20 TeV giving a potentially serious restriction on this type of model.  

The examples we have considered in the main body of this paper avoid this problem as the interactions determining the freeze-in dynamics come from renormalisable or super-renormalisbale operators.   Moreover, if the excitations of the ``Higgs" field that converts the HDO into a $d\leq 4$ operator approach  the intermediate scale, as might be expected if the associated VEVs are at this scale, then no significant restriction on $T_R$ arises.  Overall, there are many examples of non-renormalisable operators and their significance in each case is somewhat model dependent.

It needs to be emphasised that similar issues occur with freeze-out in the presence of non-renormalisable operators involving a thermally decoupled particle that is either stable or long enough lived that it decays after the freeze-out of the would be dark matter.  Since extensions of the SM involving GUT-unification, gravity, or extra dimensions often have such particles and couplings, freeze-out theories of dark matter genesis similarly suffer from restrictions on $T_R$ when viewed in this broader context.  The most well studied example involves the gravitino where the yield $Y_{\tilde{G}} \sim 10^{-12} (T_R/ 10^{10} \mbox{GeV})$ also approximately increases linearly with $T_R$, leading to important limits on the reheat temperature for weak-scale gravitino mass, though there are also other examples of particles interacting via non-renormalisable operators with analogous limits on $T_R$, see e.g. \cite{ewimp}.

\subsection{Freezing-in a FIMP sector}

Throughout this paper we have focussed on the case of one FIMP species. We note here some of the potential extensions of this idea by considering more complicated FIMP sectors.  For example, consider the case where we have one FIMP species, $X$, coupled to a thermal bath via some renormalisable operator.  It is through this operator that an abundance of $X$ is frozen in. 
In addition to this suppose $X$ has further interactions with some other states, $X_i$, which have even weaker or no direct interactions with the thermal bath.  The strength of the interactions between the $X_i$s and $X$ could be moderate or even large. Consequently, once an abundance of $X$s are frozen-in these extra interactions could bring the $X_i$s and $X$ particles into partial thermal equilibrium and within this sector a secondary freeze-out process could take place leaving a relic abundance of the lightest state in this sector.  If the lightest state is stable it can form the dark matter but if it is not it decays back into an LSP or similar in the visible sector.
This process of secondary freeze-out can substantially alter the final dark matter abundance, changing the observable phenomenology associated to the FIMP.

A further variation arises when there is more than one FIMP particle, for example we can imagine that there are several moduli (perhaps with different masses) that couple via some operator to particles in the thermal bath. The couplings of these moduli are feeble and so each could have an abundance frozen-in. If the moduli have different coupling strengths (or masses) the abundances of each moduli will be different.  Depending on the mass spectrum these frozen-in moduli could be the dark matter or decay to the real dark matter particle.

\section{Conclusions}\label{conclusions}

The nature and origin of the dark matter in the universe is unknown.  While there are many theories of DM,
there are rather few cosmological production mechanisms, and even fewer that can be subjected to precision tests.
For example, the non-thermal coherent oscillation of a scalar field, such as an axion, always involves an unknown 
initial field amplitude.  Thermal mechanisms that are independent of initial conditions are particularly interesting,
and the decoupling of a heavy particle species from the thermal bath has received enormous attention.  Such 
decoupling can occur when the particle is relativistic, as with the neutrinos, or non-relativisitic, as with baryons.  
The former leads to hot dark matter, so that the latter case of thermal freeze-out has been widely studied for DM.

At first sight these appear to be the only two ways of thermally producing DM without any sensitivity to 
initial conditions.  If a massive particle of the thermal bath is to survive the early universe with a 
significant abundance it must decouple, and this can either happen when the particle is relativistic or 
non-relativistic.  We have argued that there is one other possibility:  the massive particle may have a negligible
initial abundance and may be produced by collisions or decays of particles in the thermal bath.   
This is only a new possibility if the massive particle does not reach thermal equilibrium, and hence requires 
that it is a FIMP, interacting very feebly with the bath.  Since the FIMP abundance is heading towards 
thermal equilibrium we call this production mechanism ``freeze-in."  

Freeze-in requires a very small coupling $\lambda$ between the FIMP and the bath  
\begin{equation}
\lambda^2 \sim \frac{T_{eq}}{M_{Pl}} \,\, \frac{m_{FI}}{m_{DM}}
\label{eq:lambda2}
\end{equation}
where $T_{eq}$ is the temperature of matter-radiation equality,
$m_{FI}$ is the mass of the heaviest particle involved in the freeze-in reaction and $m_{DM}$ 
is the mass of the DM particle.  The mechanism works for a very wide range of FIMP masses: 
to be produced it must be lighter than the reheat temperature after inflation, and for DM to be 
cold it must be heavier than a keV.   Indeed the range of theories involving FIMP freeze-in is so large 
that in this paper we have concentrated on FIMPs with a mass of order the weak scale, such as
moduli or right-handed sneutrinos.

It is not surprising that freeze-out is so popular: it presumably occurred with baryons (although with 
the added complication of a chemical potential) and it works with dimensionless coupling parameters of 
order unity.  Indeed, with couplings of order unity, if the only mass scale in the 
annihilation cross section is that of the DM particle, then 
this mass is predicted to be of order the weak scale.  On the other hand freeze-in of a FIMP requires a 
special situation.  The freeze-in reaction must involve a small dimensionless coupling parameter,
eq.~(\ref{eq:lambda2}); furthermore, UV sensitivity may reappear unless higher dimensional operators are 
suppressed or the reheat temperature after inflation is quite low.  However, small couplings for both 
renormalisable and higher dimensional operators occur very easily, for example by approximate 
symmetries or small wavefunction overlaps in higher dimensions.  Most importantly, precisely 
because FIMPs necessarily have very small couplings to the bath, eq~(\ref{eq:lambda2}), they offer 
the prospect of exotic signals of dark matter generation.  

If the FIMP is the lightest particle carrying the DM stabilising symmetry it is the DM, so that the lightest 
observable sector particle (LOSP) with this symmetry is unstable and decays to the FIMP with a rate 
suppressed by $\lambda^2$.  Alternatively it may be the LOSP that is DM, 
and in this case it is the FIMP that has a very long lifetime.   The signals associated with the 
freeze-in mechanism all revolve around the long lifetime of decays between the FIMP and LOSP.  
In particular, we explored the signals that arise when the 
FIMP, LOSP and freeze-in masses are all of order the weak scale. 

In many theories a small coupling may arise from a parametric form that is linear in the weak scale $v$,
$\lambda \sim v/M_{Pl}$.    In fact, in many models the Planck scale will be replaced by the string or
compactification scale, as occurs for certain moduli, providing a better explanation
for the order of magnitude of $\lambda$.   The temperature of matter-radiation equality from DM produced
by the freeze-in mechanism then has the same parametric form as in the WIMP case: $T_{eq} \sim v^2 / M_{Pl}$.
This not only explains why 
freeze-in can yield the observed abundance, but suggests that two mechanisms for DM production -- 
FIMP freeze-in and LOSP freeze-out -- yield broadly comparable abundances.

The cosmological signals of the FIMP arise from the late decays between the FIMP and the LOSP and,
with masses of order the weak scale, the parametric form of the lifetime in the simplest theories is
\begin{equation}
\tau \, \sim \, \frac{M_{Pl}}{T_{eq}\, v} \, \sim \, 1 \, \mbox{sec.}
\label{eq:tau}
\end{equation}
A precise computation shows that if the FIMP is frozen-in via two body decays or inverse decays 
that involve the LOSP, then the lifetime is closer to $10^{-2}$ seconds, as shown in eq.~(\ref{eq:tauB1}).
However, this should be viewed as a lower bound on the lifetime of decays between the FIMP and 
LOSP.  Further suppression of the decay rate occurs if the freeze-in process involves bath particles 
other than the LOSP or if the decays are to three or more particles.  Thus the 
lifetime may well be in the region where the decays can alter the nuclear abundances produced during 
BBN, possibly solving the $^7Li$ problem, and may even be long enough to 
allow a component of the DM to be warm.

If FIMP freeze-in dominates over LOSP freeze-out, then the LOSP annihilation cross section is larger 
than with conventional freeze-out.  This leads to an important astrophysical signal of LOSP DM: 
the DM annihilation in the galactic halo is boosted relative to the conventional case, leading to
enhanced rates for indirect detection signals of photons, leptons and anti-protons.   

Finally, FIMP dark matter yields important new collider signals.  The exotic particles of the 
LOSP sector can be produced at the LHC and will rapidly decay to the LOSP, which has a long lifetime, 
bounded from below by about $10^{-2}$ sec.  If the LOSP is charged or coloured it has a significant 
probability of being stopped in the detector, so that its decays can be observed.  Studying the production 
and decay of the LOSP at the LHC could directly probe the freeze-in reaction that created the FIMP dark 
matter in the early universe.

\section*{Acknowledgements}
We thank Asimina Arvanitaki, Savas Dimopoulos, Sergei Dubovsky, Piyush Kumar and Scott Watson for discussions. 
KJ, JMR, and SMW gratefully thank the Berkeley Center for Theoretical Physics for their warm hospitality during the course of this work.  LH, KJ and SMW gratefully acknowledge hospitality from the Dalitz Institute for Fundamental Physics, Oxford University and for visitor support from the European ERC Advanced Grant 228169-BSMOXFORD. JMR is partially supported by the EC network 6th Framework Programme Research and Training Network Quest for Unification (MRTN-CT-2004-503369), by the EU FP6 Marie Curie Research and Training Network UniverseNet (MPRN-CT-2006-035863),  by the STFC (UK), and by a Royal Society Wolfson Award. SMW thanks the Higher Education Funding Council for England and the STFC (UK), for financial support under the SEPNet Initiative. The work of LH was supported in part by the National Science Foundation under grant PHY-0457315 and  in part by the Director, Office of Science, Office of High Energy and Nuclear Physics, of the US Department of Energy under Contract DE-AC02-05CH11231.

\end{document}